\documentclass[pre,aps,showpacs,nofootinbib,superscriptaddress]{revtex4}
\usepackage{graphicx}
\usepackage{natbib}
\usepackage{bm}
\usepackage{amssymb}
\usepackage{amsmath}
\usepackage{euscript}
\usepackage{esint}
\usepackage{prelim2e}

\begin{document}

\title{Stability analysis of confined V-flames. I. Analytical treatment \\
of the high-velocity limit}

\author{Hazem El-Rabii}
\author{Guy Joulin}

\affiliation{%
Laboratoire de Combustion et de D\'etonique, CNRS/ENSMA, 1 av.
Cl\'ement Ader, 86961
Futuroscope, Poitiers, France}%

\author{Kirill A. Kazakov}
\affiliation{%
Department of Theoretical Physics, Physics Faculty, Moscow State
University, 119899, Moscow, Russian Federation}

\begin{abstract}
The problem of linear stability of confined V-flames with arbitrary gas expansion is addressed. Using the on-shell description of flame dynamics, a general equation governing propagation of disturbances of an anchored flame is obtained. This
equation is solved analytically for V-flames in high-velocity channel streams. It is demonstrated that dynamics of flame disturbances in this case is controlled by the memory effects associated with vorticity generated by the curved front. The perturbation growth rate spectrum is determined, and explicit analytic expressions for the eigenfunctions are given. It is found that the piecewise linear V-structure is unstable for all values of the gas expansion coefficient.

\end{abstract}
 \maketitle
% \narrowtext
% \unitlength=1pt

\section{Introduction}

Among the various types of premixed flame propagation problems, anchored flames
hold a special place. On the one hand, such flames are
relatively easy to realize experimentally; on the other, they look simple
enough for theoretical investigation, because they admit several important simplifications. For instance, open flames anchored by means of a thin rod are often observed to have rectilinear wings (unconfined V-flames). Homogeneity of the upstream flow, adopted usually as the natural approximation compatible with this piecewise linear flame-front structure, often conveys the impression that the problem is easily solvable analytically. It thus represents an excellent laboratory for testing
our understanding of premixed flame dynamics.

Despite these promising circumstances there is an apparent lack of
theoretical results on V-flame properties. The reason is that these
flames are not as simple as they seem. A closer inspection of the
flow structure of the idealized V-configuration reveals that this
pattern is singular: the pressure field turns out to diverge
logarithmically near the tip of the flame-front (and also at
infinity along the front, in the case of unconfined V-flames). This is
a sign of incompleteness of the idealized picture, which means that
the system anchoring the flame must be explicitly included into
consideration. This essential complication necessitates the introduction
of a specific inner scale in the problem (in addition to the
cutoff wavelength and the channel width), thereby raising the question as to the influence
of this new scale on the whole basic pattern. The initial problem is
thus naturally divided into two parts: 1) modeling of the anchoring
system; this primarily is a stationary analysis, aimed at inferring
properties of the system needed to generate a presumed flame pattern, and
2) investigation of the flame dynamics, which first and foremost is a
stability analysis of the anchored flame; an important issue in this
analysis is its model-dependence, {\it i.e.,} the extent to which its results
depend on particularities of the anchoring system.

The purpose of the present paper is to carry out an analysis of the
above-mentioned issues, in the case of a confined V-flame anchored in a high-velocity gas stream. It will be shown that the problem admits a full theoretical investigation in this important particular case, and that its results are model-independent in the above sense. It should be mentioned that in contrast to unconfined anchored flames, flames anchored in channels do not exhibit an acute linear structure, although the piecewise linear front with a uniform upstream flow is still a solution of the governing equations. Experiments show that deviations from linearity occur not only in the small regions near the anchor and the channel walls, but all along the front \citep{scurlock1948,progreport1949,zeldo1985}. This suggests that the simplest configuration is possibly unstable in the confined case. The results of our work fully confirm this conjecture.

In our investigation, we use the on-shell description of flames developed in Refs.~\citep{kazakov1,kazakov2,jerk1,jerk2}. The integro-differential equations derived therein provide a non-perturbative
description of spontaneous flame dynamics in the most general form,
{\it i.e.,} they apply to flames with arbitrary gas expansion and
arbitrary jump conditions across the flame front. The main advantage
of using these equations is that they are {\it closed}, in the sense
that they involve only quantities defined {\it at} the flame front. This
allows one to avoid explicit solving of the flow equations in the
bulk, which is the stumbling block of conventional analysis. This approach will be shown to be extendable to the case of anchored
flames in a simple and natural way.

The paper is organized as follows. Section \ref{formulation} serves
to set up the general framework of the on-shell flame description. In Sec.~\ref{spontan}, we formulate the
problem and recall the main results of Refs.~\citep{kazakov1,kazakov2,jerk1,jerk2}.
Extension of these results to the case of anchored flames is
described in Sec.~\ref{anchored}. An analysis of the anchoring system
impact on the flame structure, carried out in
Sec.~\ref{rodinfluence}, is used in Sec.~\ref{boundaryconds} to
identify boundary conditions for the linearized equation describing the
propagation of disturbances. This equation is derived, in a form
suitable for the subsequent analysis, in Sec.~\ref{linearizedeq}, then
solved in Sec.~\ref{onshelldynamics}. An
important step here is the evaluation of rotational contribution,
presented in Sec.~\ref{roteval}. The resulting equation is analyzed
in the high-velocity limit in Sec.~\ref{larges}, which allows
considerable simplifications. In particular, an asymptotic expansion
of the main integral operator $\EuScript{H}$ is constructed in
Sec.~\ref{hexpansion}. Finally, analytic solutions of the linearized
problem are found in Sec.~\ref{analytsol}, and studied in detail in
Sec.~\ref{solstructure}. Section \ref{conclusions} contains
concluding remarks and prospects for future work. The paper has two
appendices, one of which contains a consistency check for the
calculations performed, and the other describes in detail transition to the case of vanishingly small anchor dimensions within the large-slope expansion.

\section{Preliminaries}\label{formulation}

\subsection{Spontaneous flame dynamics on-shell}\label{spontan}

Consider a 2D-flame propagating in a channel of constant width $b,$
filled with an initially quiescent uniform ideal gas. Let
the Cartesian coordinates $(x,y)$ be chosen so that the channel
walls are at $x=0,b,$ and $y = - \infty$ is in the fresh gas. These
coordinates will be measured in units of the channel
width\footnote{However, we keep track of $b$ throughout
Sec.~\ref{formulation}.}, while fluid velocity, $\bm{v} = (w,u),$ in
units of the velocity of a plane flame front relative to the fresh
gas, $U_f.$ Finally, the fluid density will be normalized by the
fresh gas density, $\theta> 1$ denoting its ratio to that of burnt
gas. We assume that the flame pattern is continued to the whole
$x$-axis in the usual way using the ideal boundary conditions at the
channel walls:
\begin{eqnarray}\label{bcond}
f' =0, \quad w=0 \quad {\rm for} \quad x = 0,b\,.
\end{eqnarray}
\noindent Then the on-shell value, $(w_-,u_-)$, of fresh-gas
velocity ({\it i.e.,} its value at the flame front considered as a
gasdynamic discontinuity), and the flame front position, $f(x,t),$
satisfy the following complex integro-differential equation
\citep{jerk1,jerk2}
\begin{eqnarray}\label{generalc1}&&
2\left(\omega_-\right)' + \left(1 +
i\hat{\EuScript{H}}\right)\left\{[\omega] - \frac{i}{4}
\int\limits_{-\infty}^{+\infty}d\tilde{x}(i\partial_y - \partial_x)
\int\limits_{\tau_-}^{\tau_+}d\tau M(\tilde{x},t-\tau)\right\}' =
0\,,
\end{eqnarray}
\noindent where $\omega = u + iw$ is the complex gas velocity,
$[\omega]$ its jump across the flame front, $\partial_x \equiv
\partial/\partial x,$ $\partial_y \equiv
\partial/\partial y,$ and the prime denotes
differentiation with respect to $x$ (in the last term on the left,
the argument $y$ is understood to be set equal to $f(x,t)$ after
partial spatial differentiation, but before the $x$-differentiation
denoted by the prime; we recall that the improper
$\tilde{x}$-integral in this term is understood as an analytic
continuation of the corresponding regularized expression, see
Ref.~\citep{jerk2} for details). The memory kernel $M$ has the form
$M(\tilde{x},t) \equiv
N(\tilde{x},t)\bar{v}^n_+(\tilde{x},t)\sigma_+(\tilde{x},t),$ where
$N = \sqrt{1 + \left(f'\right)^2}\,,$ $\bar{v}^n_+ = \bar{v}_{i+}
n_i$ is the normal burnt gas velocity relative to the flame front,
$n_i$ denoting the unit vector normal to the front ($\bm{n}$ points
towards the burnt gas), and $\sigma_+$ is the on-shell value of
vorticity produced by the curved front. The memory kernel is
integrated over any path in the complex time-plane, connecting the
points $$\tau_{\pm} = \frac{r}{\bar{v}_+}\left(\Omega \pm i\sqrt{1 -
\Omega^2}\right)\,, \quad \Omega \equiv
\frac{(\bm{r}\cdot\bar{\bm{v}}_+)}{r\bar{v}_+}\,,$$ where
$$\bar{\bm{v}}_+ = (w_+,\bar{u}_+)\,, \quad \bar{u}_+(x,t) \equiv
u_+(x,t) - \frac{\partial f(x,t)}{\partial t}$$ is the on-shell
burnt gas velocity relative to the front, and $\bm{r}$ is the
radius-vector drawn from the point $(\tilde{x},f(\tilde{x},t))$ at
the front to the observation point $(x,y).$ Finally, the action of
the operator $\hat{\EuScript{H}}$ on an arbitrary function $a(x)$ is
defined by
\begin{eqnarray}\label{hcurved}
\left(\hat{\EuScript{H}}a\right)(x) = \frac{1 + i
f'(x,t)}{\pi}~\fint\limits_{-\infty}^{+\infty}
d\tilde{x}~\frac{a(\tilde{x})}{\tilde{x} - x + i[f(\tilde{x},t) -
f(x,t)]}\,,
\end{eqnarray}
\noindent where the slash denotes the principal value of the
integral. For a $2b$-periodic function $a(x)$ [{\it i.e.,} $a(x+2b) =
a(x)$], summing explicitly the integrand with the help of the
formula
$$\sum\limits_{k=-\infty}^{+\infty} \frac{1}{2bk + z} =
\frac{\pi}{2b}\cot\left(\frac{\pi z}{2b}\right),$$ the right hand
side of (\ref{hcurved}) can be rewritten as an integral over the
channel width
\begin{eqnarray}\label{hcurvedf}
\left(\hat{\EuScript{H}}a\right)(x) = \frac{1 + i
f'(x,t)}{2b}~\fint\limits_{-b}^{+b}
d\tilde{x}~a(\tilde{x})\cot\left\{\frac{\pi}{2b}(\tilde{x} - x +
i[f(\tilde{x},t) - f(x,t)])\right\}\,.
\end{eqnarray}
\noindent We recall also that the value of vorticity at the front
and the normal velocity of the burnt gas, entering the function
$M(\tilde{x},t-\tau),$ as well as the velocity jumps at the front,
are all known functionals of on-shell fresh gas velocity
\citep{matalon,pelce}. For zero-thickness flame fronts one has
\begin{eqnarray}\label{jumps}
\bar{v}^n_+ &=& \theta\,, \quad [u] = \frac{\theta - 1}{N}\ ,
\quad [w] = - f'\frac{\theta - 1}{N}\,, \\
\sigma_+ &=& - \frac{\theta - 1}{\theta N}\left[\frac{Dw_-}{Dt} +
f' \frac{Du_-}{Dt} + \frac{1}{N}\frac{Df'}{Dt}
\right]\,,\label{vorticity}
\end{eqnarray}
\noindent where
$$\frac{D}{Dt} \equiv \frac{\partial}{\partial t} + \left(w_- +
\frac{f'}{N}\right)\frac{\partial}{\partial x}\,.$$ Together with
the evolution equation
\begin{eqnarray}\label{evolutiongen}
(\bar{\bm{v}}_-\cdot \bm{n}) = 1\,,
\end{eqnarray}
\noindent the complex Eq.~(\ref{generalc1}) constitutes a closed
system of three equations for the three functions $w_-(x,t),$
$u_-(x,t)$ and $f(x,t).$

\subsection{On-shell description of anchored flames}\label{anchored}

As derived, Eq.~(\ref{generalc1}) describes only spontaneous flame
evolutions. However, the anchoring system is not difficult to
incorporate into the framework of the on-shell description. This can
be done as follows. Consider the simplest and most commonly used in
practice type of the anchoring system -- a metal rod placed
somewhere within the channel. From the mathematical point of view,
the presence of the rod can be described as a singularity of the
complex velocity, $\omega = u + iw,$ considered as an analytical
function of the complex variable $z=x+iy.$ Namely, suppose that the
original field, $\omega_0(z),$ is superimposed with the complex
velocity, $\omega^d(z),$ describing a dipole located at the point
$(x_0, y_0)$:
\begin{eqnarray}\label{diapole}
\omega_0(z) + \frac{d}{(z-z_0)^2} \equiv \omega(z)\,,
\end{eqnarray}
\noindent where $z_0 = x_0 +i y_0,$ and $d=d_1+id_2$ is a complex
constant determining strength of the dipole as well as its
orientation. For sufficiently small $|d|,$ perturbation of the main
flow is noticeable only in a small vicinity of the dipole. Since
$\omega_0(z)$ is analytical at $z=z_0,$ one has
$$\omega_0(z) = \omega_0(z_0) + O(|z-z_0|),$$ and hence, the complex
velocity near the dipole can be written approximately as
\begin{eqnarray}\label{diapole1}
\omega(z) \approx \omega_0(z_0) + \frac{d}{(z-z_0)^2}\,.
\end{eqnarray}
\noindent The form of the stream lines is given by
$${\rm Re}\left\{\omega_0(z_0)(z - z_0) - \frac{d}{z-z_0}\right\} = {\rm const}\,,$$
or $$u_0(x-x_0) - w_0(y-y_0) - \frac{d_1(x-x_0) +
d_2(y-y_0)}{(x-x_0)^2 + (y-y_0)^2} = {\rm const}\,,
$$ where $u_0,w_0$ are the real and imaginary parts of
$\omega_0(z_0).$ It is seen that if we choose $d_1 = u_0R^2,$ $d_2 =
- w_0R^2,$ with $R$ arbitrary real, then the stream-line family
contains a circle of radius $R,$ centered at the point $(x_0,y_0).$
Thus, adding the term $\omega^d(z) = \omega^*_0(z_0)R^2/(z-z_0)^2$
to the velocity field $\omega_0(z)$ describes perturbation of the
given flow by a cylindrical rod of radius $R,$ centered at $z_0.$ To
take into account non-uniformity of the main flow near the rod, and
to describe more general rod profiles, it will be necessary to
superpose several dipoles located within the rod area, and to
include higher-order multipoles into consideration.

To obtain generalization of Eq.~(\ref{generalc1}) to the case of
anchored flames, we recall that this equation is a consequence of
the following relations:
\begin{eqnarray}\label{chup}
\left(1 - i \hat{\EuScript{H}}\right)\left(\omega_-\right)' = 0\,,
\\ \label{chdown}
\left(1 + i \hat{\EuScript{H}}\right)\left(\omega^p_+\right)' = 0\,,
\\ \label{vortonshell}\omega^v_+ = \frac{i}{4}
\int\limits_{-\infty}^{+\infty}d\tilde{x}(i\partial_y - \partial_x)
\int\limits_{\tau_-}^{\tau_+}d\tau M(\tilde{x},t-\tau)
\\ \label{identpv} \omega^p_+ = - \omega^v_+ + \omega_- +
[\omega]\,,
\end{eqnarray}
\noindent where $[\omega]$ denotes the jump of the complex velocity
across the flame front, $[\omega] = \omega(x,f(x,t)+0) -
\omega(x,f(x,t)-0).$ Equations (\ref{chup}), (\ref{chdown}) express
analyticity and boundedness of the complex velocity upstream, and
its potential component downstream \citep{kazakov1,kazakov2},
Eq.~(\ref{vortonshell}) is the on-shell expression of the rotational
component \citep{jerk2}, while Eq.~(\ref{identpv}) is an obvious
identity. As we have just seen, the presence of the rod violates
analyticity of the complex velocity, so that either of
Eqs.~(\ref{chup}), (\ref{chdown}) is no longer valid, depending on
whether the rod is placed up- or downstream. In the former case,
Eq.~(\ref{chup}) is satisfied by $\omega_0(z) = \omega(z) -
\omega^d(z),$ because it {\it is} analytical upstream and bounded.
On the other hand, since $\omega^d(z)$ does not have singularities
downstream and is bounded there, it satisfies Eq.~(\ref{chdown}).
Thus,
\begin{eqnarray}
\left(1 - i \hat{\EuScript{H}}\right)\left(\omega_- -
\omega^d_-\right)' = 0\,,\nonumber
\\ \label{chddown}
\left(1 + i \hat{\EuScript{H}}\right)\left(\omega^d_+\right)' =
0\,.\nonumber
\end{eqnarray}
\noindent Since $\omega^d_- = \omega^d_+,$ we see that
Eq.~(\ref{chup}) is replaced in this case by
\begin{eqnarray}\label{chup1}
\left(1 - i \hat{\EuScript{H}}\right)\left(\omega_-\right)' =
2\left(\omega^d_-\right)'\,.
\end{eqnarray}
\noindent Accordingly, acting on Eq.~(\ref{identpv}) by the operator
$(1 + i \hat{\EuScript{H}}),$ we obtain the following equation
\begin{eqnarray}\label{generalc1a}&&
2\left(\omega_-\right)' + \left(1 +
i\hat{\EuScript{H}}\right)\left\{[\omega] - \frac{i}{4}
\int\limits_{-\infty}^{+\infty}d\tilde{x}(i\partial_y - \partial_x)
\int\limits_{\tau_-}^{\tau_+}d\tau M(\tilde{x},t-\tau)\right\}' =
2\left(\omega^d_-\right)'\,,
\end{eqnarray}
\noindent which is the sought extension of Eq.~(\ref{generalc1}) to
the case of anchored flames. In the case of the rod located
downstream, similar considerations show that Eq.~(\ref{chdown}) must
be replaced by the following
\begin{eqnarray}
\left(1 + i \hat{\EuScript{H}}\right)\left(\omega^p_+ -
\omega^d_+\right)' = 0\,,\nonumber
\\ \label{chdownd}
\left(1 - i \hat{\EuScript{H}}\right)\left(\omega^d_+\right)' = 0\,.
\end{eqnarray}
\noindent It is not difficult to verify that the resulting equation
for $\omega_-$ in this case has exactly the same form
(\ref{generalc1a}).

\subsection{Influence of anchoring system on V-flame
structure}\label{rodinfluence}

As was mentioned in introduction, the necessity of explicit
inclusion of the anchoring system into consideration raises the
question as to what extent this system affects global properties of
V-flames. Let us now show that as long as linear dimensions of the
rod are small compared to the channel width, so that the flame front
can be considered piecewise linear, influence of the rod on the
flame structure is local, in the sense that it is confined to a
small region near the rod. We recall, first of all, that the
relative value of the velocity disturbance caused by a dipole
modeling the rod is proportional to $R^2/(x^2 + y^2)$ (for
simplicity, the dipole is assumed to be at the origin). Hence, under
the assumption $R\ll b,$ this disturbance is indeed negligible for
the most part of the channel, except a small region $(x,y\sim R)$
near the rod. This simple reasoning is not yet sufficient to prove
our statement, because it only demonstrates the locality of, so to speak,
direct rod influence on the flow structure. In such an essentially
nonlocal problem as deflagration, we also have to look for
possible indirect consequences of this influence, related to the
fact that the presence of the rod ultimately determines the basic
flame pattern. The on-shell description is particularly convenient
for this purpose, as it explicitly reveals the nonlocal structure of
the governing equations.

For the rest of the paper, flames will be considered in the
reference frame attached to the rod (the above-given formulation is
invariant under transitions between different reference frames).
Accordingly, the fresh-gas velocity at infinity will be denoted $U$:
$$u(x,y=-\infty,t) = U (> 0).$$

\begin{figure}
\centering
\includegraphics[width=.5\textwidth]{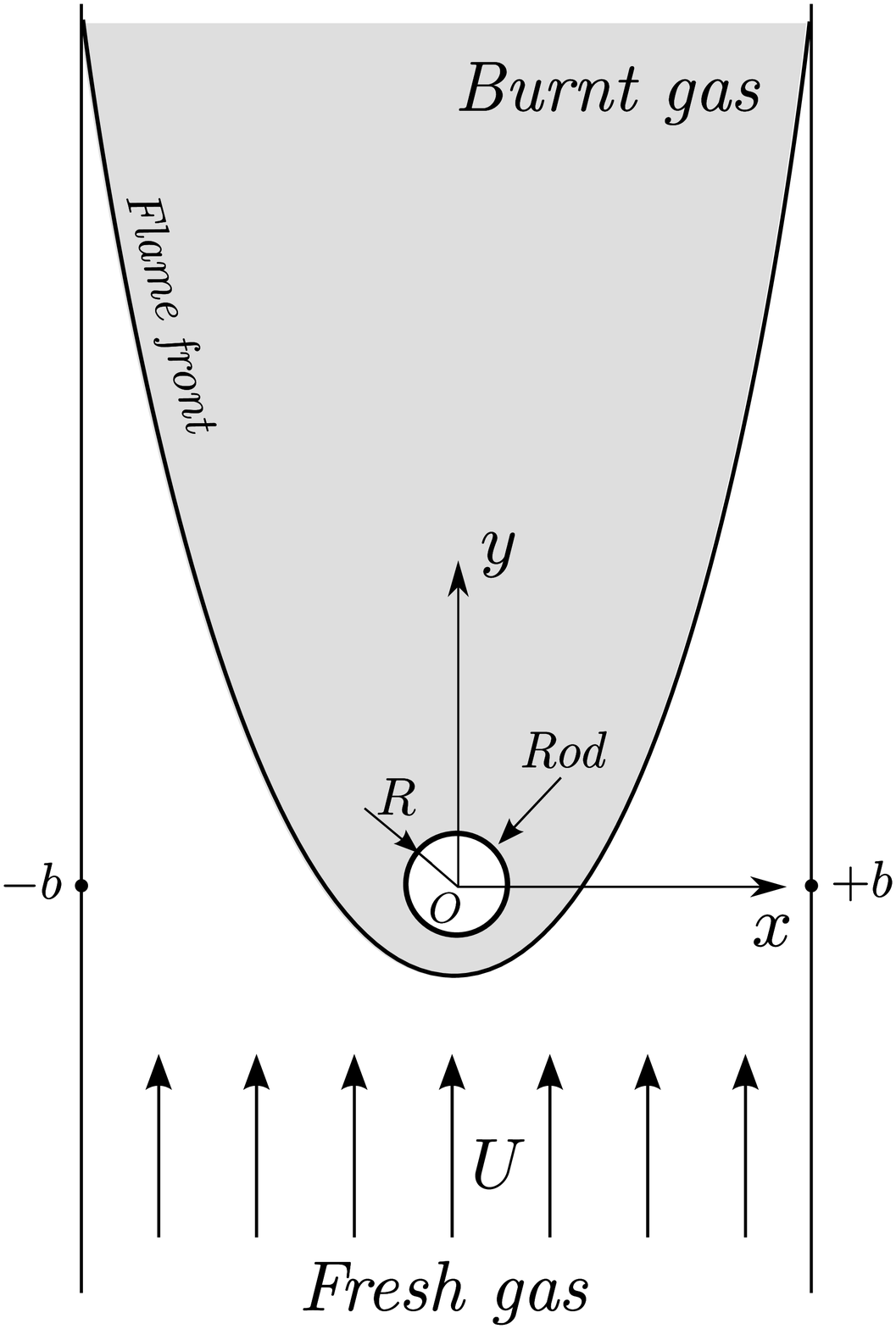}
\caption{}\label{fig1}
\end{figure}

We will assume in what follows that the anchoring system is
stationary, {\it i.e.,} its properties do not change with time. This
means that these properties can be inferred from the steady-state
V-flame structure. To this end, we note that the stationary version
of Eq.~(\ref{generalc1a}) reads (here we are in the rest frame of
the flame-front, so the over-bar in the notation of velocity is omitted)
\begin{eqnarray}\label{generalc1st}&&
2\left(\omega_-\right)' + \left(1 +
i\hat{\EuScript{H}}\right)\left\{[\omega]' -
\frac{Nv^n_+\sigma_+\omega_+}{v^2_+} \right\} =
2\left(\omega^d_-\right)',
\end{eqnarray}
\noindent which follows directly from the fact that
Eq.~(\ref{generalc1}) reduces in this case to the stationary
equation derived in \citep{kazakov1,kazakov2}. In regions where the flame-front
slope is constant and the upstream flow is homogeneous, the first
term on the left as well as the expression in the curly brackets
vanish, because velocity jumps are constant there, $\omega_- = {\rm
const},$ and vorticity is not produced. This expression is only
non-zero in a vicinity of the rod where all the quantities involved
vary rapidly. It is this rapid variation that is a possible source
of indirect influence of the rod on the global flame structure.
Indeed, for $R/b\to 0,$ both terms in the curly brackets have a
$\delta$-functional character. If the $\delta$-singularity were not
canceled in their sum, then upon the action of the
$\EuScript{H}$-operator it would give rise to an expression which is
non-zero everywhere in the channel. However, we have just seen that
the right hand side of Eq.~(\ref{generalc1st}) vanishes outside of
small region around the rod. Therefore, in order that this equation
be satisfied, the $\delta$-contributions must cancel. To be more
specific, let us assume that the rod is located downstream (which is
normally the case in actual experiments), as shown in
Fig.~\ref{fig1}. Then, using Eqs.~(\ref{chup}), (\ref{chdownd}), and
the identity $\omega_- + [\omega] = \omega_+,$
Eq.~(\ref{generalc1st}) can be conveniently rewritten as
\begin{eqnarray}&&\label{generalc1st1}
\left(1 + i\hat{\EuScript{H}}\right)\left\{\left(\omega_+\right)' -
\frac{Nv^n_+\sigma_+\omega_+}{v^2_+} \right\} =
\left(1+i\hat{\EuScript{H}}\right)(\omega^d_+)'\,.
\end{eqnarray}
\noindent There are two types of $\delta$-like contributions on the
left hand side of this equation, corresponding to the real and
imaginary parts of the expression in the curly brackets. Since the
real part is even under $x\to -x,$ its $x$-derivative is
odd. Hence, the corresponding singularity is generally
proportional to $\delta'(x),$ and can be compensated by
appropriately choosing the coefficient $d$ in the dipole term on the
right hand side. Indeed, the on-shell value of a dipole $\omega^a(z)
= a/(z - a)^2,$ considered in the limit $a\to 0,$ possesses all
characteristic properties of the $\delta$-function: $\omega_+(0) =
1/a \to \infty,$ $\omega_+(x) \to 0,$ for $x\ne 0,$ and the integral
$\int_{\Delta} dx a/(x+if(x) - a)^2,$ taken over a region $\Delta
\gg a$ around $x=0,$ has a finite value (because $f(x)$ is an even function).

Things are different, however, for the imaginary part which is odd
in $x.$ In this case, the singularity is proportional to
$\delta(x);$ for zero-thickness flames, for instance, contribution
of the first term in the curly brackets to the singularity is equal
to $-2i(\theta - 1)s\delta(x)/\sqrt{1+s^2},$ where $s$ is the value
of the front slope far from $x=0,$ as is seen from
Eq.~(\ref{jumps}). Singularities of this kind\footnote{In fact, it is
the singularities $\sim\delta(x),$ with undifferentiated
$\delta$-functions, which are only important. Indeed, on dimensional
grounds, a differentiated $\delta$ should be accompanied by an extra
factor with the dimension of length; since this an ``inner''
contribution, the factor is $\sim R,$ and hence the
$\delta'(x)$-terms can be neglected in comparison with $\delta(x)$
in the limit $R\to 0.$ Another way to see this is to recall that the
parameter $a$ in the dipole $\omega^a(z) = a/(z - a)^2$ is $O(R),$
while the strength of the dipole modeling the rod, $|d| = O(R^2),$
as we saw in Sec.~\ref{anchored}. Hence, $\omega^a(z)$ must be
accompanied by a factor $O(R).$} cannot be compensated by any local
field $\omega^d(z).$ Thus, we arrive at the conclusion that the
assumption of piecewise linear front structure implies the absence
of terms proportional to $\delta(x)$ on the left hand side of
Eq.~(\ref{generalc1st1}), {\it i.e.,} that the contribution of the
first term in the curly brackets is canceled by that of the second
term. This requirement can be written in the following integral form
\begin{eqnarray}\label{dcond}
\int_{\Delta} dx \left\{\left(w_+\right)' -
\frac{Nv^n_+\sigma_+w_+}{v^2_+}\right\} = o(1) \quad {\rm for} \quad
\Delta/b \to 0\,,
\end{eqnarray}
\noindent where $\Delta$: $R\ll \Delta \ll b$ is the length scale
where ``inner'' solutions ($|x|\ll b$) are to be matched  with the
``outer'' ones ($|x|\gg R$). Indeed, by virtue of Eq.~(\ref{dcond}),
the contribution of the small region near the rod to the left hand side
of Eq.~(\ref{generalc1st1}) is also small outside that region, which
is just the required absence of the $\delta$-terms.
Equation~(\ref{dcond}) thus represents a condition that selects
inner solutions compatible with the prescribed global flame
structure.

\subsection{Linearized equation for flame
perturbations}\label{linearizedeq}

In the present paper, we for are looking for possible genuine
V-flame instabilities, which would be inherent to
the V-configuration itself, and unrelated to the properties of
a specific anchoring system. We thus assume, as was already mentioned, that
this system is stationary, and the condition (\ref{dcond}) is
fulfilled. Then the equation for flame perturbation is obtained by
linearizing Eq.~(\ref{generalc1a}) around the stationary solution,
with the right hand side kept fixed. This linearized equation thus
coincides formally with that derived in Ref.~\citep{jerk1,jerk2}, but for
our present purposes another form of this equation will be more
appropriate, which avoids explicit differentiation of the memory
kernel.

First of all, since the basic pattern is stationary, time-dependence
of perturbations factorizes:
\begin{eqnarray}\label{flowperturb}
\delta f(x,t) = \tilde{f}(x)e^{\nu t}\,, \quad \delta w_-(x,t) =
\tilde{w}(x)e^{\nu t}\,, \quad \delta u_-(x,t) = \tilde{u}(x)e^{\nu
t}\,,
\end{eqnarray}
\noindent where $\nu$ is a complex constant to be found as part of the solution. Not to mix the
imaginary unit entering $\nu$ with that appearing in
Eq.~(\ref{generalc1a}), we will denote the former by $j$: $$\nu =
\nu_1 + j\nu_2,$$ where $\nu_{1,2}$ are real numbers. Accordingly,
the amplitudes $\tilde{f},\tilde{w},\tilde{u}$ are to be understood
complex with respect to $j$ (until Sec.~\ref{solstructure}, $j$ will
not appear in formulas explicitly; an example illustrating the use
of this ``double imaginary unit'' formalism is given in Appendix~A).
Next, taking into account that the basic solution is piecewise
constant, we obtain the following equation for the $x$-dependent
parts of the perturbations
\begin{eqnarray}\label{linearized1}
2\tilde{\omega}' + \left(1 +
i\hat{\EuScript{H}}\right)\left\{[\tilde{\omega}] -
\frac{1}{2}\int\limits_{-\infty}^{+\infty}d\tilde{x}\hat{M}_{\alpha}({\tilde{x}})
\tilde{\xi}_{\alpha}(\tilde{x})\frac{\omega_+}{v^2_+} \exp\left(-
\frac{\nu r}{v_+}e^{-i\phi}\right)\chi(x - \tilde{x})\right\}' =
0\,,
\end{eqnarray}
\noindent where $\{\tilde{\xi}_{\alpha}\} =
(\tilde{f},\tilde{w},\tilde{u})\,,$ $\hat{M}_{\alpha}({\tilde{x}})$
is the differential operator obtained by linearizing the function
$M(\tilde{x},t)$ around the stationary solution, and setting
$\partial/\partial t \to \nu$ afterwards; $\phi\in [-\pi,+\pi]$ is
the angle between the vectors $\bm{r},$ $\bm{v}_+,$ defined positive
if the rotation from $\bm{v}_+$ to $\bm{r}$ is clockwise, and
$\chi(x)$ is the sign function,
$$\chi(x) = \left\{
\begin{array}{cc}
+1,& x>0\,,\\
\phantom{+}0, & x = 0\,,\\
-1,&  x<0\,.
\end{array}
\right.
$$
In the form written, Eq.~(\ref{linearized1}) applies to flames with
arbitrary jump conditions at the front and arbitrary local
propagation law. However, to investigate the problem as stated in
the beginning of this paragraph, we do not need to remain at such a
general level. As mentioned earlier, we are concerned with
instabilities specific to the presumed V-pattern, so the characteristic
perturbation wavelength of interest is of the order of the channel
width which is normally much larger than the cutoff wavelength.
Hence, the curvature effects can be completely neglected in our
investigation, and the consideration be limited to the case of
zero-thickness flames. Then the linearized velocity jumps, appearing in Eq.~(\ref{linearized1}), take the form
\begin{eqnarray}\label{jumpslin}&&
[\tilde{u}(x)] = - \frac{(\theta -
1)s\chi(x)\tilde{f}'(x)}{(1+s^2)^{3/2}}\ , \quad [\tilde{w}(x)] = -
\frac{(\theta - 1)\tilde{f}'(x)}{(1+s^2)^{3/2}}\,.
\end{eqnarray}
\noindent
To linearize the memory kernel, another form of the expression (\ref{vorticity}) will be more suitable, which avoids appearance of the second spatial derivatives of the flame-front position. The point is that linearizing Eq.~(\ref{vorticity}) directly is readily seen to lead to expressions of the type $\chi(x)\delta(x)$ which are not well-defined in the sense of distributions. To resolve this ambiguity, we first rewrite Eq.~(\ref{evolutiongen}) as
$$u_- -\frac{\partial f}{\partial t} - f'w_- = N\,,$$ differentiate it with respect to $x,$ and use the resulting equation to eliminate $f''$ from the right hand side of Eq.~(\ref{vorticity}). The memory kernel thus becomes
\begin{eqnarray}\label{memorykernel}
M = -(\theta-1)\left[\frac{\partial w_-}{\partial t} + f'\frac{\partial u_-}{\partial t} - u_-'\frac{\partial f}{\partial t} + w_-w_-'+u_-u_-'\right]\,.
\end{eqnarray}
\noindent The right hand side of this expression now involves only first derivatives of continuous functions. It should be emphasized that this trick works only in the outer region where effects due to the finite flame-front thickness are negligible. If they are not,  $f''$ appears already in the undifferentiated evolution equation. Linearization of Eq.~(\ref{memorykernel}) yields
\begin{eqnarray}\label{mlin}&&
\hat{M}_{\alpha}\tilde{\xi}_{\alpha}(x) = - (\theta - 1)\left[\nu\tilde{w}(x) +
s\chi(x)\nu\tilde{u}(x) + \sqrt{1+s^2}\tilde{u}'(x) \right]\,.
\end{eqnarray}
\noindent Finally, the linearized evolution equation reads
\begin{eqnarray}\label{evolutionlin}&&
\tilde{u}(x) - s\chi(x)\tilde{w}(x) = \nu\tilde{f}(x) +
\frac{s\chi(x)\tilde{f}'(x)}{\sqrt{1+s^2}}\,.
\end{eqnarray}
\noindent

\subsubsection{Boundary conditions}\label{boundaryconds}

We consider symmetrical basic V-patterns, so that the
flame-anchoring rod is located in the middle of the channel. For
simplicity, flame disturbances also will be assumed symmetrical
under reflection with respect to the $y$-axis. In these
circumstances, it is convenient, without changing notation, to
consider a double-width channel occupying the strip $-b\leqslant x
\leqslant +b,$ with the rod being at the origin $x=y=0,$ and the
$y$-axis playing the role of the symmetry plane of the flame.
Accordingly, boundary conditions of the exact problem, {\it i.e.,}
the problem we started with, read $w=f'=0$ for $x=\pm b.$ They are
used, in particular, for the periodic continuation of the flame
pattern, mentioned in Sec.~\ref{spontan}. After the initial problem
is divided into the inner and outer ones, these conditions naturally
remain pertaining to the former. A question thus arises as to the
boundary conditions relevant to the outer problem.

Evidently, the requirement of a vanishing front slope at the
walls is now irrelevant. Indeed, it is not satisfied already by the
basic steady V-configuration defined to have the constant slope,
$|f'|=s,$ everywhere. The front flattening takes place in a thin
boundary layer near the walls, characterized by large values of
$f''.$ On the other hand, the $x$-derivative of the $w$-component
does not have to be large in this region, as can be seen from the
fact that the condition $w=0,$ being the universal boundary
condition for the ideal fluid, is compatible with any prescribed
front configuration. Indeed, the linearized Euler equations for the
fresh-gas read
\begin{eqnarray}\label{eulerw}
\nu\delta{w} + U\frac{\partial\delta{w}}{\partial y} &=& -
\frac{\partial \delta{p}}{\partial x}\,, \\
\label{euleru} \nu\delta{u} + U\frac{\partial\delta{u}}{\partial y}
&=& - \frac{\partial \delta{p}}{\partial y}\,,
\end{eqnarray}
\noindent where it is taken into account that for the steady-state
V-flame, $u = U,$ $w=0.$ Using the continuity equation in
Eq.~(\ref{euleru}), multiplying it by $f',$ subtracting from
Eq.~(\ref{eulerw}), and going over on shell one obtains an equation
for $\delta w_-$:
$$\left(\frac{\partial \delta w}{\partial x}\right)_- +
f'\left(\frac{\partial \delta w}{\partial y}\right)_- =
\left(\delta{w}_-\right)' = \frac{1}{U}\left\{\nu\left(\delta{u}_- -
f'\delta w_-\right) + \left(\frac{\partial \delta{p}}{\partial y} -
f'\frac{\partial \delta{p}}{\partial x}\right)_-\right\}\,,$$ or
\begin{eqnarray}\label{deltawderiv}
\frac{d\delta{w}_-}{dl} = \frac{1}{U}
\left\{\nu(\delta\bm{v}_-,\bm{n}) + \left(\frac{\partial \delta
p}{\partial n}\right)_-\right\}\,,
\end{eqnarray}
\noindent where $l$ is the front length counted off from the channel
wall, and $\bm{n}$ is, as usual, the unit vector normal to the
front. As was already mentioned, the front flattens in a thin
boundary layer near the walls. From the standpoint of the outer
problem we are concerned with, a finite change of an on-shell
variable across the layer is seen as a finite jump of that quantity
at the channel wall. Accordingly, its derivative with respect to $l$
contains a term proportional to $\delta(l).$ Now suppose that this
is the case for $\delta w_-(l).$ Then it follows from
Eq.~(\ref{deltawderiv}) that
$$\nu(\delta\bm{v}_-,\bm{n}) + \left(\frac{\partial \delta
p}{\partial n}\right)_- = c\delta(l) + \cdots,$$ where $c$ is a
constant, and ``$\cdots$'' denote regular terms. Since the velocity
jump is finite, this means that
$$\left(\frac{\partial \delta p}{\partial n}\right)_- =
c\delta(l) + \cdots.$$ This relation can be integrated by noting
that differentiation of pressure in the direction {\it normal} to
the front cannot produce a $\delta$-singularity {\it along} the
front, and hence,
$$\delta p = \EuScript{C}\delta(l) + \cdots,$$ where $\EuScript{C}$
is such that $(\partial
\EuScript{C}/\partial n)_- = c.$ But pressure is only allowed to
have a finite jump at the wall, therefore, $\EuScript{C}$ must vanish upstream. Meanwhile, in the absence of obstacles at the wall, the outer solution is regular on-shell, and hence the function $\EuScript{C}$ is differentiable not only in the near upstream region (as required by its definition), but also at the flame front. Under such circumstances, the requirement $\EuScript{C}=0$ upstream entails vanishing of its derivative at the front, {\it i.e.,} $c=0.$ Thus, $\delta{w}_-$ is in fact continuous at the wall, so its vanishing remains a boundary condition of the outer problem: $\delta
w_-(\pm b) = 0,$ or
\begin{eqnarray}\label{boundarycond1}&&
\tilde{w}(\pm b) = 0\,.
\end{eqnarray}
\noindent

The reasoning just given does not apply at $x=0,$ because of the presence of the rod. Nevertheless, $\tilde{w}(0)$ must also vanish, as a consequence of our assumption that the anchoring system is stationary. To see this, let us consider the procedure of matching the inner and outer solutions in more detail. Take the $x$-component of the fresh-gas velocity. For gas elements moving near the $y$-axis, this component is zero everywhere except a small vicinity of the rod. More precisely, $w$ induced by the rod is $O(U)$ for $\sqrt{x^2 + y^2}\equiv \rho \sim R,$ and rapidly decreases with distance. At distances $\rho \sim R_0,$ where $R\ll R_0 \ll b,$ the inner solution describing the flow near the rod is matched with the outer solution we are interested in. In the steady case, matching at the flame front assigns $w_-$ a definite value, say~$w_0,$ which is generally nonzero. This value plays the role of a boundary condition for the steady flow, defining thereby the basic pattern. Now, since the properties of the rod are assumed stationary, in particular, unaffected by perturbations of the outer solution, so is the flow near the rod. Therefore, matching of the inner solution with the outer one will give $w$ the same value $w_0.$ In other words, $\delta w_-|_{\rho\sim R_0} = \delta w_0 = 0\,,$ which in the limit $R, R_0 \to 0$ yields
\begin{eqnarray}\label{boundarycond2}&&
\tilde{w}(0) = 0\,.
\end{eqnarray}
\noindent By the same reasoning,
\begin{eqnarray}\label{boundarycond2u}&&
\tilde{u}(0) = 0\,.
\end{eqnarray}
\noindent Finally, the remaining condition replacing $\tilde{f}'=0$ is
\begin{eqnarray}\label{boundarycond3}&&
\tilde{f}(0) = 0\,.
\end{eqnarray}
\noindent It follows directly from the fact that we consider the rod
dimension as vanishingly small compared to the channel width.
Indeed, the linear dimension of the flame tip as well as its
separation from the rod are both $\sim R.$ Hence, $f(x\sim R) \sim
R,$ which in the limit $R\to 0$ gives Eq.~(\ref{boundarycond3}).
This condition means that the flame is not torn off from the rod by
the perturbation.

\section{On-shell dynamics of V-flame
perturbations}\label{onshelldynamics}

\subsection{Evaluation of the rotational
contribution}\label{roteval}

\begin{figure}
\centering
\includegraphics[width=.8\textwidth]{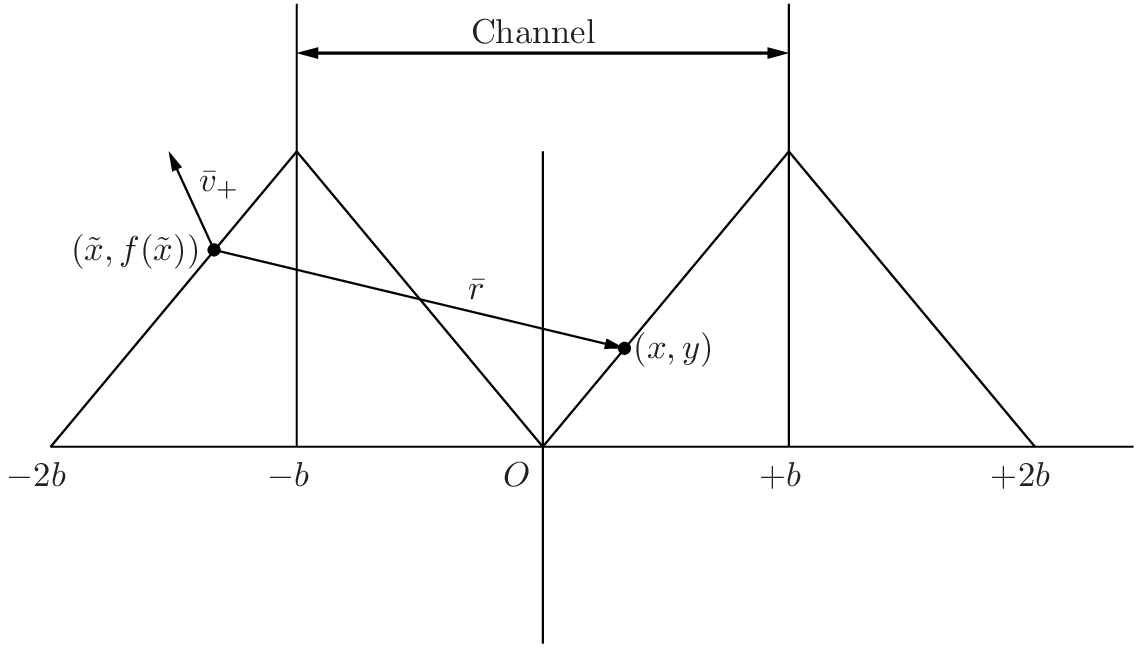}
\caption{}\label{fig2}
\end{figure}

In order to study evolution of the V-flame disturbances using
Eq.~(\ref{linearized1}), we have to evaluate the improper
$\tilde{x}$-integral appearing in the curly brackets. We recall that
this integral is understood as an analytic continuation of the
regularized integral
\begin{eqnarray}\label{integral}
\int\limits_{-\infty}^{+\infty}d\tilde{x}e^{-\mu |\tilde{x} -
x|}\hat{M}_{\alpha}({\tilde{x}})
\tilde{\xi}_{\alpha}(\tilde{x})\frac{\omega_+}{v^2_+} \exp\left(-
\frac{\nu r}{v_+}e^{-i\phi}\right)\chi(x - \tilde{x})\,,
\end{eqnarray}
\noindent to the limit $\mu \to 0^+.$ To simplify the calculation, we note that
$$rv_+e^{-i\phi} = -i(z - \tilde{z})\omega_+\,, \quad \tilde{z} = \tilde{x} + is|\tilde{x}|\,.$$ Indeed, one has $|z - \tilde{z}| = r,$ $|\omega_+| = v_+,$ while according to the definition of the angle $\phi$ it is equal to the phase difference of the complex functions $w_+ + iu_+ = i\omega^*_+$ and $(z - \tilde{z}).$
We need to consider two different situations corresponding to the integration
variable running over the negative- or positive-slope part of the
flame front (see Fig.~\ref{fig2}). Assuming that the observation
point $x\in [0,1],$ one has, in the case $\tilde{x}\in [- 2n,-2n +
1],$ $n \in Z,$
$$z - \tilde{z} = (x - \eta + 2n) + i s(x - \eta)\,,$$ where $[0,1] \ni \eta = \tilde{x} + 2n\,.$ Similarly, in
the case $\tilde{x}\in [-1 - 2n,-2n],$
$$z - \tilde{z} = (x - \eta + 2n) + is(x + \eta)\,,$$ where $[-1,0] \ni \eta = \tilde{x} + 2n\,.$ In effect, the exponent in the integrand of (\ref{integral}) takes the form
$$\exp\left(- \frac{\nu r}{v_+}e^{-i\phi}\right) = \left\{
\begin{array}{cc}
\exp\left( - \displaystyle\frac{\nu\omega^*_0}{|\omega_0|^2}\left[-
i(x -
\eta + 2n) + s(x - \eta)\right]\right)\,,& \eta \in [0, + 1] \,,\\
\exp\left( - \displaystyle\frac{\nu\omega_0}{|\omega_0|^2}\left[-
i(x - \eta + 2n) + s(x + \eta)\right]\right)\,,&  \eta\in [-1, 0]\,,
\end{array}
\right. $$ where $$\omega_0 = U + (\theta - 1)\frac{1 +
is}{\sqrt{1+s^2}}\,.$$ Furthermore, the regularizing factor $e^{-\mu |\tilde{x} -
x|}$ may be replaced by $e^{-\mu 2|n|},$ because $(x - \eta)$ is finite. Next, the $\tilde{x}$-integral taken over $(-\infty,+\infty)$ can be represented as an integral over $\eta \in [-1,+1]$ of the integrand summed over all $n.$ Since the functions $\hat{M}_{\alpha}({\tilde{x}}),$ $\omega_+(\tilde{x})$ are periodic by construction, we need to sum the following series
$$I(\mu) = \sum\limits_{n=-\infty}^{+\infty} \exp\left\{
2ni\varkappa - 2|n|\mu\right\}\chi(x - \eta + 2n)\,,$$ where
$$\varkappa = \left\{
\begin{array}{cc}
\nu/\omega_0\,,& \eta \in [0, + 1] \,,\\
\nu/\omega^*_0\,,&  \eta\in [-1, 0]\,.
\end{array}
\right.$$ Taking into account that $|x - \eta|\leqslant 2,$ one has
\begin{eqnarray}
I(\mu) &=& \chi(x - \eta) + \sum\limits_{n=0}^{+\infty} \exp\left\{
2n(i\varkappa - \mu)\right\} - \sum\limits_{n=0}^{+\infty}
\exp\left\{ 2n(-i\varkappa - \mu)\right\} \nonumber \\
&=& \chi(x - \eta) + \left[ 1 - \exp\{2(i\varkappa -
\mu)\}\right]^{-1} - \left[ 1 - \exp\{2(-i\varkappa -
\mu)\}\right]^{-1}\,.
\end{eqnarray}
\noindent Since the initial improper $\tilde{x}$-integral is reduced
to an integral over a finite domain, its analytic continuation to
$\mu = 0^+$ amounts to that of the function $I(\mu),$ which is
$$I(0^+) = \chi(x - \eta) + \left[ 1 - \exp\{2i\varkappa \}\right]^{-1}
- \left[ 1 - \exp\{-2i\varkappa\}\right]^{-1} = \chi(x - \eta) +
i\cot\varkappa\,.$$

All these formulas were derived for $x \in [0,+1].$ From these, the
corresponding expressions for $x \in [-1,0]$ can readily be obtained
by noting that the integral (\ref{integral}) is invariant under the
combined operations of inversion $x \to - x,$ and complex
conjugation. This rule can be deduced directly from the explicit
formulas (\ref{jumpslin}), (\ref{mlin}), taking the various parity
properties of the flow variables into account, yet it is in fact a general
property of the formalism, unrelated to the specific approximations made. In
what follows, we will denote this combined operation as $(x\to
-x)^*.$ It should be kept in mind that the complex conjugation here
is understood with respect to the imaginary unit $i,$ but not to
$j$:
$$i^* = - i, \quad j^* = j.$$

Putting all these results into Eq.~(\ref{linearized1}), we thus
arrive at the following linearized equation governing evolution of
the flame disturbances
\begin{eqnarray}\label{linearized2}&&
2\tilde{\omega}' + \frac{\theta - 1}{2}\left(1 +
i\hat{\EuScript{H}}\right)\left\{\frac{e^{i\varkappa(x +
is|x|)}}{\omega_0} \int\limits_{0}^{+1}d\eta \left[\nu\tilde{w}(\eta) +
s\nu\tilde{u}(\eta) + \sqrt{1+s^2}\tilde{u}'(\eta) \right]\right.\nonumber\\&& \left.\times e^{-i\varkappa(1 + is)\eta}
\left[i\cot\varkappa + \chi(x - \eta)\right] - \frac{i +
s\chi(x)}{(1+s^2)^{3/2}}\tilde{f}'(x) + (x\to -x)^*
\phantom{\int\limits_{0}^{+1}}\hspace{-0.5cm}\right\}' = 0\,,
\end{eqnarray}
\noindent where the symbol $(x\to -x)^*$ refers to the whole
expression written out explicitly in the curly brackets. As a useful check of the
calculations performed, it is verified in Appendix A that in the
particular case $s=0$ this equation reproduces the well-known
Darrieus-Landau dispersion relation \citep{darrieus,landau} for the
perturbation growth rate.

\subsection{The high-velocity limit}\label{larges}

In its general form, Eq.~(\ref{linearized2}) can presumably be
solved only numerically. It turns out, however, that it is amenable to a full
theoretical analysis in the case when the velocity of the incoming
fresh-gas flow is high: $$U \gg 1\,.$$ Being opposite to that of
classical analysis \citep{darrieus,landau,siv1,sivclav}, this limit is
of considerable interest both from practical and theoretical points
of view, as it represents the situation where propagation of the
flame disturbances is strongly affected by the basic flow. We will
see that the nonlocal interaction of flame perturbations with the
background takes a new form which is principally different from that
encountered in the conventional weak-nonlinearity analysis. Also,
dependence of solutions on the gas expansion coefficient becomes
quite intricate, having nothing in common with that found in the
small-gas-expansion approximation.

\subsubsection{Large-$s$ expansion of the
$\EuScript{H}$-operator}\label{hexpansion}

We start discussion of the high-velocity limit by deriving an
approximate expression for the $\EuScript{H}$-operator appearing in
Eq.~(\ref{linearized2}). There, it is defined at the unperturbed
front, $f(x) = s|x|,$
\begin{eqnarray}\label{hcurvedf1}
\left(\hat{\EuScript{H}}a\right)(x) = \frac{1 + i
s\chi(x)}{2}~\fint\limits_{-1}^{+1}
d\tilde{x}~a(\tilde{x})\cot\left\{\frac{\pi}{2}\left[\tilde{x} - x +
is(|\tilde{x}| - |x|)\right]\right\}\,.
\end{eqnarray}
\noindent By virtue of the relation
$$U = \sqrt{1+s^2}\,,$$ large values of $U$ imply that the front
slope is also large, so the argument of cotangent in
Eq.~(\ref{hcurvedf1}) has a large imaginary part for almost all
values of the integration variable, in which case one has
\begin{eqnarray}\label{approxcot}
\cot\left\{\frac{\pi}{2}\left[\tilde{x} - x + is(|\tilde{x}| -
|x|)\right]\right\} \approx - i\chi(|\tilde{x}| - |x|)\,.
\end{eqnarray}
\noindent This approximation is valid for all $\tilde{x}$ except two
small regions near $\tilde{x} = \pm|x|\,.$ More precisely, taking
into account that, for real $a_{1,2},$ $$\cot(a_1+ia_2) =
-i\,\frac{e^{(a_2 - ia_1)} + e^{-(a_2 - ia_1)}}{e^{(a_2 - ia_1)} -
e^{-(a_2 - ia_1)}} = -i\chi(a_2) + O\left(e^{-2|a_2|}\right)\,,$$ we
see that Eq.~(\ref{approxcot}) holds true, {\it with an exponential
accuracy}, everywhere except $$\tilde{x}: |\tilde{x}| \in (|x| -
\delta, |x| + \delta),$$ where $\delta = O(1/s).$

\begin{figure}
\centering
\includegraphics[width=.8\textwidth]{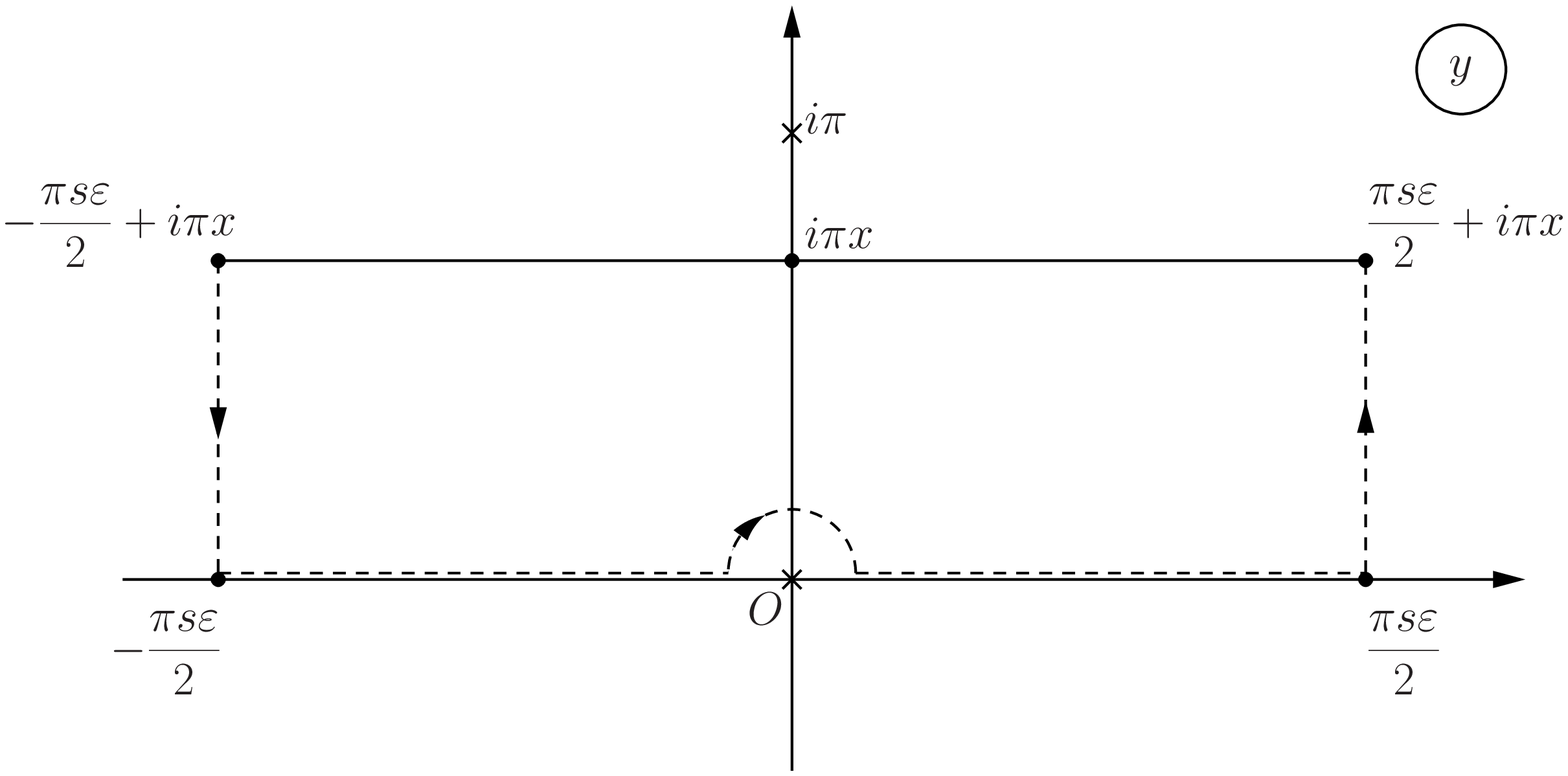}
\caption{}\label{fig3}
\end{figure}

To develop an asymptotic expansion of $\hat{\EuScript{H}}$ in powers
of $1/s$ for $s \gg 1,$ let us choose a real $\varepsilon > 0$
satisfying
\begin{eqnarray}\label{epsiloncond}
\varepsilon \ll 1\,, \quad s\varepsilon \gg 1\,.
\end{eqnarray}
\noindent Then the integral in Eq.~(\ref{hcurvedf1}) can be
rewritten as
\begin{eqnarray}\label{intred}&&
\fint\limits_{-1}^{+1}
d\tilde{x}~a(\tilde{x})\cot\left\{\frac{\pi}{2}\left[\tilde{x} - x +
is(|\tilde{x}| - x)\right]\right\} = -
i\left[\int\limits_{-1}^{-x-\varepsilon} +
\int\limits_{-x+\varepsilon}^{0} + \int\limits_{0}^{x-\varepsilon} +
\int\limits_{x+\varepsilon}^{+1}\right]
d\tilde{x}~a(\tilde{x})\chi(|\tilde{x}| - x) \nonumber\\&& +
\left[\int\limits_{-x-\varepsilon}^{-x+\varepsilon} +
\fint\limits_{x-\varepsilon}^{x+\varepsilon}\right]d\tilde{x}~a(\tilde{x})\cot\left\{\frac{\pi}{2}\left[\tilde{x}
- x + is(|\tilde{x}| - x)\right]\right\}\,,
\end{eqnarray}
\noindent where we assumed that $x>0,$ for definiteness. Notice that
in the last term on the right hand side of Eq.~(\ref{intred}), only
one of the two integrals is defined in the principal value sense. As
such, it is proportional to the derivative of $a(x).$ It is not
difficult to verify that contributions of this kind give rise to
terms of the order $1/s^2.$ Below, we will need $\hat{\EuScript{H}}$
expanded up to $O(1)$-terms, so the principal-sense integral can be
neglected. The other integral can be evaluated as follows, within
this accuracy,
\begin{eqnarray}&&
\int\limits_{-x-\varepsilon}^{-x+\varepsilon}d\tilde{x}~a(\tilde{x})
\cot\left\{\frac{\pi}{2}\left[\tilde{x} - x + is(|\tilde{x}| -
x)\right]\right\} = - i a(-x)
\int\limits_{-\varepsilon}^{+\varepsilon}d\tilde{x}~ \coth \left\{\frac{\pi s}{2}\tilde{x} + \pi ix\right\} \nonumber\\&& =
- i a(-x)\frac{2}{\pi s} \int\limits_{-\pi s\varepsilon/2 + \pi i
x}^{+\pi s\varepsilon/2 + \pi i x}d y~\coth y\,.\label{yintegral}
\end{eqnarray}
\noindent By virtue of the conditions (\ref{epsiloncond}), the
$y$-integral can be calculated, with exponential accuracy,
using the contour deformation shown in Fig.~\ref{fig3}
\begin{eqnarray}&&
\int\limits_{-\pi s\varepsilon/2 + \pi i x}^{+\pi s\varepsilon/2 +
\pi i x}d y~\coth y = \left[-\int\limits_{-\pi s\varepsilon/2 +
\pi i x}^{-\pi s\varepsilon/2} + \int\limits_{+\pi s\varepsilon/2
}^{+\pi s\varepsilon/2 + \pi i x}\right]d y + \fint\limits_{-\pi
s\varepsilon/2 }^{+\pi s\varepsilon/2}d y~\coth y - i\pi = \pi
i(2x - 1)\,.\nonumber
\end{eqnarray}
\noindent On the other hand, replacing cotangent by the sign
function gives zero within the same accuracy
\begin{eqnarray}&&
\int\limits_{-x-\varepsilon}^{-x+\varepsilon}d\tilde{x}~a(\tilde{x})
\chi(|\tilde{x}| - x) = a(-x)
\int\limits_{-\varepsilon}^{+\varepsilon}d\tilde{x}~\chi(\tilde{x})
= 0\,.\nonumber
\end{eqnarray}
\noindent Using these results in Eq.~(\ref{intred}), and then
substituting it in Eq.~(\ref{hcurvedf1}) gives finally
\begin{eqnarray}\label{hcurvedf2}
\left(\hat{\EuScript{H}}a\right)(x) = (s\chi(x) -
i)~\int\limits_{0}^{+1} d\tilde{x}~\frac{a(\tilde{x}) +
a(-\tilde{x})}{2}\chi(\tilde{x} - |x|)+ia(-x)(2|x| - 1) +
O\left(\frac{1}{s}\right),
\end{eqnarray}
\noindent where the symmetry of the operator $i\hat{\EuScript{H}}$
under $(x\to -x)^*$ was taken into account to dismiss the condition
$x>0.$ As a special case of this formula, let us consider the action of $\hat{\EuScript{H}}$ on a derivative. If $a(x)$ satisfies
$a(0^+) = a(0^-),$ $a(+1) = a(-1),$ then integrating by parts in Eq.~(\ref{hcurvedf2}) readily gives
\begin{eqnarray}\label{hcurvedf3}
\left(\hat{\EuScript{H}}a'\right)(x) = (s\chi(x) -
i)\left\{- a(|x|)+ a(-|x|)\right\} + ia'(-x)(2|x| - 1) +
O\left(\frac{1}{s}\right),
\end{eqnarray}
\noindent where the prime now denotes the derivative of the function
with respect to its argument, $a'(y) = da(y)/dy.$ It turns out that this formula holds true even if the function $a(x)$ does not satisfy the above conditions of periodicity and continuity at the origin. This is proved in Appendix B.

To conclude this section, some comments concerning the structure of
the expression (\ref{hcurvedf2}) are in order. First of all, it is
seen that the result of the action of $\hat{\EuScript{H}}$ depends
essentially on parity properties of the function $a(x),$ namely,
$\hat{\EuScript{H}}a = O(s),$ if $a(x)$ is even, and
$\hat{\EuScript{H}}a = O(1),$ if it is odd. Next, the appearance of a
term proportional to $a(-x)$ encodes a peculiar
interaction between the points $x$ and $-x,$ which is natural taking
into account that the front wings get close to each other in the limit $s\to \infty.$ Finally, it should be noted
that although the identity $\hat{\EuScript{H}}\circ
\hat{\EuScript{H}} = - 1$ is valid whatever the shape of the
flame-front, in particular, in the large-$s$ limit, it cannot be
verified using the expression on the right of Eq.~(\ref{hcurvedf2}),
already because of the composition of its leading term with the
undetermined remainder $O(s)\circ O(1/s) = O(1).$

\subsubsection{Equation for the $x$-component of velocity.
Relative order of the flow perturbations}\label{dconds}

The results of the previous section allow us to a obtain a simple relation between components of the perturbed on-shell velocity. We use the following equation
\begin{eqnarray}\label{chuplin}
\left(1 - i \hat{\EuScript{H}}\right)\tilde{\omega}' = 0\,,
\end{eqnarray}
\noindent which is obtained acting by $(1 - i
\hat{\EuScript{H}})$ on Eq.~(\ref{linearized2}), and using the
identity $\hat{\EuScript{H}}\circ \hat{\EuScript{H}} = - 1$; it can
be derived also directly by linearizing Eq.~(\ref{chup}).
Applying the formula (\ref{hcurvedf3}) yields
\begin{eqnarray}\label{chuplinr}
\tilde{w}(x) = \frac{1 - |x|}{s}\tilde{u}'(x)\,.
\end{eqnarray}
\noindent
We see that the boundary condition (\ref{boundarycond1}) is met explicitly, while setting $x=0$ and using (\ref{boundarycond2}) leads to a new condition
\begin{eqnarray}\label{boundarycond5}
\tilde{u}'(0) = 0 \,.
\end{eqnarray}
\noindent It will be shown in the next section that this condition is also satisfied automatically by the solutions of Eq.~(\ref{linearized2}).

Next, we use Eq.~(\ref{chuplinr}) to determine the
relative order of the flow perturbations within the large-$s$
expansion. It is convenient to assume that $\tilde{u} = O(1).$ It
follows then from Eq.~(\ref{chuplinr}) that $\tilde{w} = O(1/s),$
while using these in the linearized evolution equation
(\ref{evolutionlin}) tells us that $\tilde{f} = O(1).$ Applying these estimates to Eq.~(\ref{linearized2}) shows immediately that the term $(i + s\chi)\tilde{f}'/(1+s^2)^{3/2}$ in the curly brackets can be omitted. Since the $\eta$-integral is explicitly continuous at $x=0,$ so is the expression in the curly brackets, as was to be shown.

In connection with this result, it is worth mentioning that the
term $(i + s\chi)\tilde{f}'/(1+s^2)^{3/2}$ represents the linearized velocity jumps which define the Frankel potential-flow equation \citep{frankel1990}. That this contribution is negligible means the evolution of disturbances in the case under consideration is essentially rotational, and cannot be described within the
potential-flow model.

\subsection{Analytical solution of the linearized equation in the high-velocity  limit}\label{analytsol}

We are now in position to proceed to analytical solving of
Eq.~(\ref{linearized2}) in the case of high stream-velocity. Although
the following calculation is a straightforward application of the
formulas derived in the preceding section, it is somewhat lengthy.
We give it in considerable detail because some of its points are
definitely worth to be mentioned.

\subsubsection{Derivation of the integro-differential
equation}\label{derivation}

To begin with, it is convenient to rewrite Eq.~(\ref{linearized2})
as
\begin{eqnarray}\label{linearized3}&&
2\tilde{\omega}' + \frac{\theta - 1}{2}\left(1 +
i\hat{\EuScript{H}}\right)E' = 0\,, \\ E(x) \equiv &&
\frac{e^{i\varkappa(x + is|x|)}}{\omega_0} \int\limits_{0}^{+1}d\eta
\left[\nu\tilde{w}(\eta) +
s\nu\tilde{u}(\eta) + \sqrt{1+s^2}\tilde{u}'(\eta) \right]\nonumber\\ && \times e^{-i\varkappa(1 + is)\eta}
\left[i\cot\varkappa + \chi(x - \eta)\right] + (x\to -x)^*\,.\label{efunction}
\end{eqnarray}
\noindent The order of the leading contribution to the left hand
side of Eq.~(\ref{linearized3}) can be read off from its first term,
$\tilde{\omega}'.$ According to the estimates of the previous
section, it is $O(1),$ and is contained in the real part of the
equation. To extract the relevant contribution from the integral
term, we recall that the action of $\hat{\EuScript{H}}$ on odd and
even functions gives rise to terms of the order $O(s)$ and $O(1),$
respectively. Furthermore, taking into account that $\omega_0 =
O(s),$ and hence $\varkappa = O(1/s),$ one sees that $E(x) = O(s).$
Therefore, according to the naive power counting the integral term
is formally $O(s^2).$ However, there is actually no discrepancy in
the orders of the two terms, because the $O(s)$-contribution to
$E(x)$ turns out to be imaginary {\it even}, and thus cancels with
its counterpart from $(x\to - x)^*.$ Yet, the formal estimate means
that expanding imaginary part of $E(x),$ one must generally keep
terms up to the second relative order in $1/s.$ With this in mind,
we write
\begin{eqnarray}\label{expand1}
\omega_0 = s\left[1 + \frac{i(\theta - 1)}{s} + \frac{\theta -
1/2}{s^2}\right]\,,
\end{eqnarray}
\noindent and then
\begin{eqnarray}\label{expand2}
e^{i\varkappa(x - 1) - \varkappa s (|x| - 1)} =&& e^{-\nu(|x| - 1)}
\left[1 + \frac{i\nu}{s}(x - 1) + \frac{i\nu(\theta - 1)}{s}(|x| -
1) \right.\nonumber \\&&\left. + \frac{\nu(\theta - 1)}{s^2}(x - 1)
+ \frac{\nu(\theta^2 - \theta + 1/2)}{s^2}(|x| - 1)\right]\,.
\end{eqnarray}
\noindent On the other hand, since in the factor $\left[\nu\tilde{w} +
s\nu\tilde{u} + \sqrt{1+s^2}\tilde{u}'\right]$ all terms are real, it can be
replaced by $s(\nu\tilde{u} + \tilde{u}'),$ with no risk of mixing orders.
Similarly, one can replace $\omega_0^{-1}\cot(\nu/\omega_0)$ by
$1/\nu,$ because the imaginary correction is $O(1/s^3).$ Also,
before expanding, it is convenient to integrate by parts the term proportional to $\tilde{u}'(\eta).$ Taking into account the boundary condition (\ref{boundarycond2u}), we thus find
\begin{eqnarray}&&
E(x) = e^{i\varkappa(x - 1) - \varkappa s (|x| - 1)}s\tilde{u}(1)
\left(\frac{i}{\nu} - \frac{1}{\omega_0}\right) +
\frac{2s}{\omega_0}\tilde{u}(x)\theta(x) \nonumber\\&& + e^{i\varkappa x -
\varkappa s |x|}i\nu\theta\int\limits_{0}^{1}d\eta
\tilde{u}(\eta)e^{-i\varkappa \eta + \varkappa s
\eta}\left(\frac{i}{\nu} + \frac{\chi(x - \eta)}{\omega_0}\right) + (x\to - x)^*, \nonumber
\end{eqnarray}
\noindent where $\theta(x)$ is the step function,
$$\theta(x) = \left\{
\begin{array}{cc}
+1,& x>0\,,\\
0,&  x\leqslant 0\,.
\end{array}
\right.
$$ Expanding further within the required accuracy with the help
of Eqs.~(\ref{expand1}), (\ref{expand2}), and omitting contributions which are real odd or imaginary even gives
\begin{eqnarray}&&
E(x) = e^{\nu (1 - |x|)}\tilde{u}(1) \left[\alpha(1 - |x|) + \frac{i
x(\alpha - \nu)}{s} \right] + 2\tilde{u}(x)\theta(x)\left(1 -
\frac{i\alpha}{s}\right)  \nonumber\\&& -
(\alpha + 1)e^{- \nu|x|}\int\limits_{0}^{1}d\eta
\tilde{u}(\eta)e^{\nu\eta}\left(1 + \frac{i\nu [x -\chi(x -
\eta)]}{s} \right) + (x\to - x)^*, \nonumber
\end{eqnarray}
\noindent where $\alpha = \theta - 1.$ It is seen that the odd
contributions are of the order $O(1/s)$ indeed, so upon the action
of $\hat{\EuScript{H}}$ they give rise to $O(1)$-terms.

Extracting the real part of Eq.~(\ref{linearized3}) with the help of the formula (\ref{hcurvedf3}) gives
\begin{eqnarray}\label{linearized3gen}
2\tilde{u}'(x) + (\theta - 1)|x|{\rm Re}E'(x) + s(\theta - 1)\chi(x){\rm Im}E(|x|) = 0\,.
\end{eqnarray}
\noindent
Since $E(x)$ is given by an integral of a piecewise continuous function [Cf. Eq.~(\ref{efunction})], it is continuous. Therefore, its imaginary part being an odd function turns into zero at the origin. Then Eq.~(\ref{linearized3gen}) tells us that its solutions satisfy the boundary condition (\ref{boundarycond5}).

Substituting the above expression for $E(x)$ in Eq.~(\ref{linearized3gen}), and introducing a new unknown function $g(x)$ according to
\begin{eqnarray}\label{ydef}
\tilde{u}(x) = g(x)e^{-\nu |x|},
\end{eqnarray}
\noindent we finally obtain the following integro-differential equation
\begin{eqnarray}&&\label{linearized4}
\left(1 + \alpha|x|\right)g'(x) - (\alpha^2 + \alpha\nu |x| +
\nu)g(x)\chi(x) + \alpha (\alpha +
1)\nu\chi(x)\int\limits_{0}^{|x|}d\eta g(\eta) \nonumber\\&& + \alpha\nu
g_1x(\alpha|x| - \alpha - 1)  = 0\,,
\end{eqnarray}
\noindent where $g_1 \equiv g(1),$ and we used
the identity
$$\int\limits_{0}^{1}d\eta g(\eta)[\chi(|x| - \eta) + 1]
= 2\int\limits_{0}^{|x|}d\eta g(\eta)\,.$$

\subsubsection{Solution of the integro-differential equation}\label{solutionint}

Up to an additive constant, Eq.~(\ref{linearized4}) is equivalent to
the following ordinary differential equation obtained by
differentiation with respect to $x$ [in view of the symmetry of this equation under $x\to -x,$ it is sufficient to consider it on the
interval $x \in (0,1)$]
\begin{eqnarray}&&\label{linearized5}
\left(1 + \alpha x\right)g''(x) - (\alpha^2 + \alpha \nu x + \nu -
\alpha)g'(x) + \alpha^2 \nu g(x) + \alpha \nu
g_1 (2\alpha x - \alpha - 1) = 0\,.
\end{eqnarray}
\noindent The general solution of this equation can be found in the form
\begin{eqnarray}&&\label{solution}
g(x) = c_1 + c_2 \nu x + h(x)\,,
\end{eqnarray}
\noindent where $c_{1,2}$ are constants, and $h(x)$ satisfies
\begin{eqnarray}&&\label{dhypergeom}
\left(1 + \alpha x\right)h''(x) - (\alpha^2 + \alpha \nu x + \nu -
\alpha)h'(x) + \alpha^2 \nu h(x) = 0\,.
\end{eqnarray}
\noindent The latter equation can be reduced to the degenerate
hypergeometric equation, and its general solution conveniently
written as
\begin{eqnarray}&&\label{solution1}
h(x) = c_3\left(x +
\frac{1}{\alpha}\right)^{\alpha}\int\limits_{\beta/\alpha}^{x +
1/\alpha}dy y^{-\alpha - 1} e^{\nu y} \,,
\end{eqnarray}
\noindent where $c_3$ and $\beta$ are new constants. A direct
substitution shows that (\ref{solution}) is a solution of
Eq.~(\ref{linearized5}), provided that the constants
$\beta,$ $c_k,$ $k=1,...,3$ satisfy
\begin{eqnarray}\label{linsyseq1}
(\alpha^2 - \alpha + \nu)\nu c_2 - \alpha^2\nu c_1 + \alpha(\alpha +
1)\nu g_1 &=& 0\,, \\ \label{linsyseq2} (\alpha -
1)\nu c_2 + 2\alpha g_1 &=& 0\,.
\end{eqnarray}
\noindent In addition to that, for (\ref{solution}) to be a
solution of the integro-differential equation~(\ref{linearized4}), the constants must be chosen so as to
guarantee vanishing of the additive constant in this equation, which
was lost upon the transition to Eq.~(\ref{linearized5}). To extract
this constant, we first of all note that
$$\left(1 + \alpha x\right)h'(x) - (\alpha^2 + \alpha\nu x +
\nu)h(x) + \alpha (\alpha + 1)\nu\int\limits_{0}^{x}d\eta h(\eta) =
c_3\left[\alpha e^{\nu/\alpha} - \nu\int\limits_{\beta}^{1}dy
y^{-\alpha - 1} e^{\nu y/\alpha}\right],$$ which can be checked by
direct computation. Then collecting the additive constants in
Eq.~(\ref{linearized4}) gives another equation for $\beta,$ $c_k$:
\begin{eqnarray}\label{linsyseq3}
\nu c_2 - (\alpha^2 + \nu)c_1 + c_3\left[\alpha e^{\nu/\alpha} -
\nu\int\limits_{\beta}^{1}dy y^{-\alpha - 1} e^{\nu y/\alpha}\right]
= 0\,.\nonumber\\
\end{eqnarray}
\noindent Finally, the boundary condition (\ref{boundarycond2u}) takes the form
\begin{eqnarray}\label{linsyseq4}
c_1 + c_3\int\limits_{\beta}^{1}dy y^{-\alpha - 1} e^{\nu y/\alpha} = 0\,.
\end{eqnarray}
\noindent Four equations (\ref{linsyseq1}) -- (\ref{linsyseq4})
constitute a closed system for the four constants $\beta,$ $c_{k}.$
In particular, the condition of consistency of this system
determines the spectrum of the perturbation growth rate $\nu.$ The
boundary value of $g(x),$ entering these equations, is expressed
through the unknowns as
\begin{eqnarray}\label{linsyseqc}
g_1 = c_1 + c_2 \nu +
c_3\int\limits_{\beta/(\alpha + 1)}^{1}dy y^{-\alpha - 1} e^{\nu
(\alpha + 1)y/\alpha} \,.
\end{eqnarray}
\noindent

\subsubsection{Reduction to an algebraic system of linear equations}

Since Eqs.~(\ref{linsyseq1}) -- (\ref{linsyseq4}) were derived from
relations linear with respect to $g(x),$ by an appropriate
redefinition of the unknowns they can be naturally rewritten as a
system of linear homogeneous equations. For this purpose, let us
introduce the following notation
\begin{eqnarray}\label{notation}
\Phi[n,\beta] &=& \int\limits_{\beta}^{1}dy y^{-\alpha - 1} e^{n
y}\,, \quad n = \frac{\nu}{\alpha}\,, \\ \label{c3c4}
c_4 &=& c_3\Phi[n,\beta]\,, \\
\Phi &=& \int\limits_{1/(\alpha + 1)}^{1}dy y^{-\alpha - 1}
e^{(\alpha + 1)ny}\,.\label{phinot}
\end{eqnarray}
\noindent It is not difficult to check that
\begin{equation*}
\int\limits_{\beta/(\alpha + 1)}^{1}dy y^{-\alpha - 1} e^{(\alpha +
1)ny} = (\alpha + 1)^{\alpha}\Phi[n,\beta] + \Phi\,.\nonumber
\end{equation*}
\noindent Using this in Eqs.~(\ref{linsyseq4}), (\ref{linsyseqc}) allows us to put them into the form that no longer involves $\beta$ explicitly:
\begin{eqnarray}
c_1 + c_4 = 0\,, \quad g_1 = c_1 + c_2 \nu + c_3\Phi +
c_4(\alpha + 1)^{\alpha}\,.\nonumber
\end{eqnarray}
\noindent On the other hand, since $g_1$ is linear with
respect to $c_k,$ $k=1,...,4,$ so are Eqs.~(\ref{linsyseq1}) --
(\ref{linsyseq4}). Therefore, taking $c_4$ as an independent unknown
instead of $\beta$ renders the system linear algebraic. Eliminating $c_4,$ we thus
obtain
\begin{eqnarray}\nonumber
c_1 \alpha + c_2 \left\{(\alpha - 1)\left[\frac{\alpha + 1}{2}n - 1\right] - n \right\} &=& 0\,, \\
\nonumber c_2(\alpha - 1)\left[\frac{\alpha + 1}{2}n - 1\right] + c_3e^n &=& 0\,,
\\ c_1\left[1 - (\alpha + 1)^{\alpha}\right] + c_2(3\alpha - 1)\frac{n}{2} + c_3\Phi &=& 0 \,.
\label{linsystem1}
\end{eqnarray}
\noindent

\subsection{Structure of the solution}\label{solstructure}

\subsubsection{The perturbation growth rate spectrum}

\begin{figure}
\centering
\includegraphics[width=.6\textwidth]{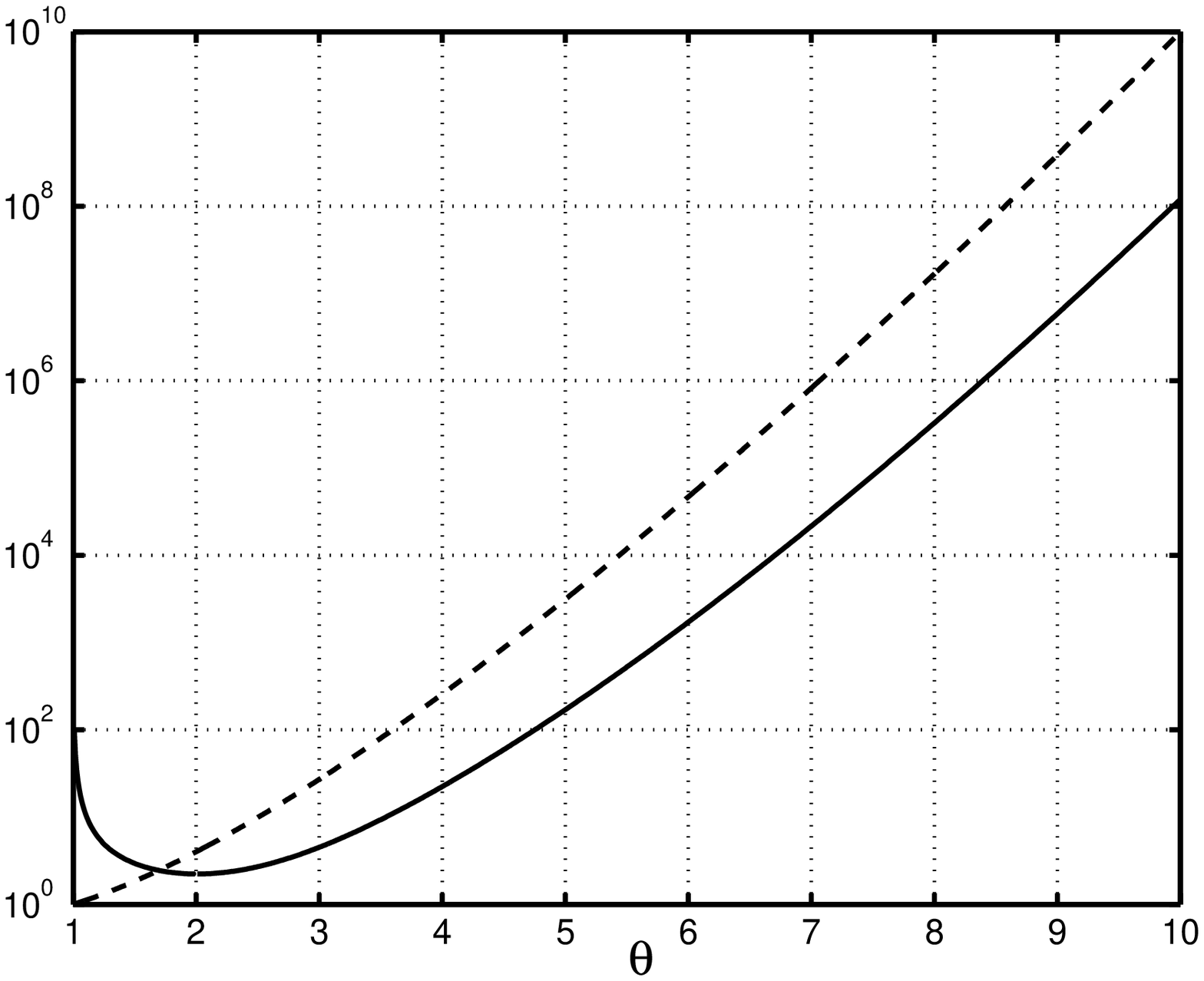}
\caption{}\label{fig4}
\end{figure}

The solvability condition for the system (\ref{linsystem1}) reads
\begin{eqnarray}\label{consist}&&
\Phi e^{-n}\alpha(\alpha - 1) \left[(\alpha + 1)n - 2\right] - n\left\{\alpha(3\alpha - 1) + \left[(\alpha + 1)^{\alpha} - 1\right](\alpha^2 - 3)\right\} \nonumber\\ && + 2(\alpha - 1)\left[(\alpha + 1)^{\alpha} - 1\right] = 0\,.
\end{eqnarray}
\noindent This equation determines the spectrum of flame
disturbances, i.e., the admissible values of the perturbation growth
rate, $\nu.$ Before looking for its numerical solutions, it is
useful to establish general features of the spectrum. For this
purpose, it is convenient to switch from $\alpha$ back to $\theta =
\alpha + 1,$ so that the definition (\ref{phinot}) takes a more
compact form
$$\Phi = \int\limits_{1/\theta}^{1}dy y^{-\theta}
e^{\theta n y}\,.$$
Integrating by parts, we can rewrite this formula for $|n|\gg 1 $ as
$$\Phi = \frac{1}{\theta n}\left\{e^{\theta n} -
\theta^{\theta}e^{n}\right\}[1 + O(1/|n|)]\,, \quad |n| \gg 1.$$
It is evident from this expression
that $\Phi \sim e^{\theta n}$ for ${\rm Re}~n \to +\infty,$ and
hence (\ref{consist}) has no solutions for such $n$'s. On the other
hand, $\Phi \sim e^{n}$ for ${\rm Re}~n \to -\infty,$ which is
compensated by the factor $e^{-n}$ in Eq.~(\ref{consist}). However,
the coefficient of the combination $\Phi e^{-n}$ as well as
the rest of the equation are polynomials in $n,$ so there are no solutions in this domain either. Thus, eigenvalues tend to be vertically aligned in the complex plane. Substituting the above asymptotic into Eq.~(\ref{consist}) yields
\begin{eqnarray}\label{consist1}
e^{\nu} = \theta^{\theta} + S(\theta)\nu\,, \quad S(\theta) \equiv
\frac{(\theta-1)(3\theta - 4)+\left[\theta^{\theta - 1} - 1\right](\theta^2-2\theta - 2)}{(\theta-1)^2(\theta - 2)}\,, \quad |\nu| \gg 1\,.
\end{eqnarray}
\noindent
Despite appearance, the function $S(\theta)$ has no pole at $\theta = 2$ (see Fig.~\ref{fig4}).

As we just mentioned, the simplified relation (\ref{consist1}) determines the spectrum in the case $|{\rm Im}~\nu | \gg 1.$ From the practical point
of view, however, we are interested in $\nu$'s whose imaginary part is not too large, so that only a finite number of eigenvalues need to be taken into account. Indeed, recalling the relation $\tilde{u}(x) = g(x)e^{-\nu x},$ the characteristic wavelength of flame perturbation with the given $\nu$ is
$$\frac{2\pi}{{\rm Im}~\nu}\,.$$ In terms of displacements along the front, $\Delta
l=s\Delta x,$ this corresponds to a wavelength $$\lambda =
\frac{2\pi s}{{\rm Im}~\nu}\,.$$ On the other hand, perturbations with wavelengths
less than the cutoff wavelength, $\lambda_c,$ are damped by the
curvature effects. The condition $\lambda \gtrsim \lambda_c$
gives, in ordinary units,
\begin{eqnarray}\label{unnumber}&&
{\rm Im}~\nu \lesssim \frac{2\pi bs}{\lambda_c}\,.
\end{eqnarray}
\noindent
For gas expansion coefficients of practical importance ($\theta = 5 \div 8$), the quantity $\theta^{\theta}$ is very large; $S(\theta)$ is also large, but smaller than $\theta^{\theta}$ by about two orders. It follows from Eq.~(\ref{consist1}) that if imaginary parts of the eigenvalues are not too large, they are close to multiples of $2\pi,$ while their real parts are approximately equal to $\theta\ln\theta,$
\begin{eqnarray}\label{grate}
\nu_m = \theta\ln\theta + j2\pi m\,, \quad m \in Z, \quad \theta\gg 1\,.
\end{eqnarray}
\noindent
This formula is useful for searching and identifying numerical solutions of the exact relation (\ref{consist}) even for smaller values of $\theta.$ Its validity as a classification scheme breaks when $S(\theta) \approx \theta^{\theta}.$ In fact, purely real solutions exist for $\theta < \theta_0 \approx 1.8.$ The corresponding modes describe aperiodic development of disturbances.

\begin{figure}
\centering
\includegraphics[width=.6\textwidth]{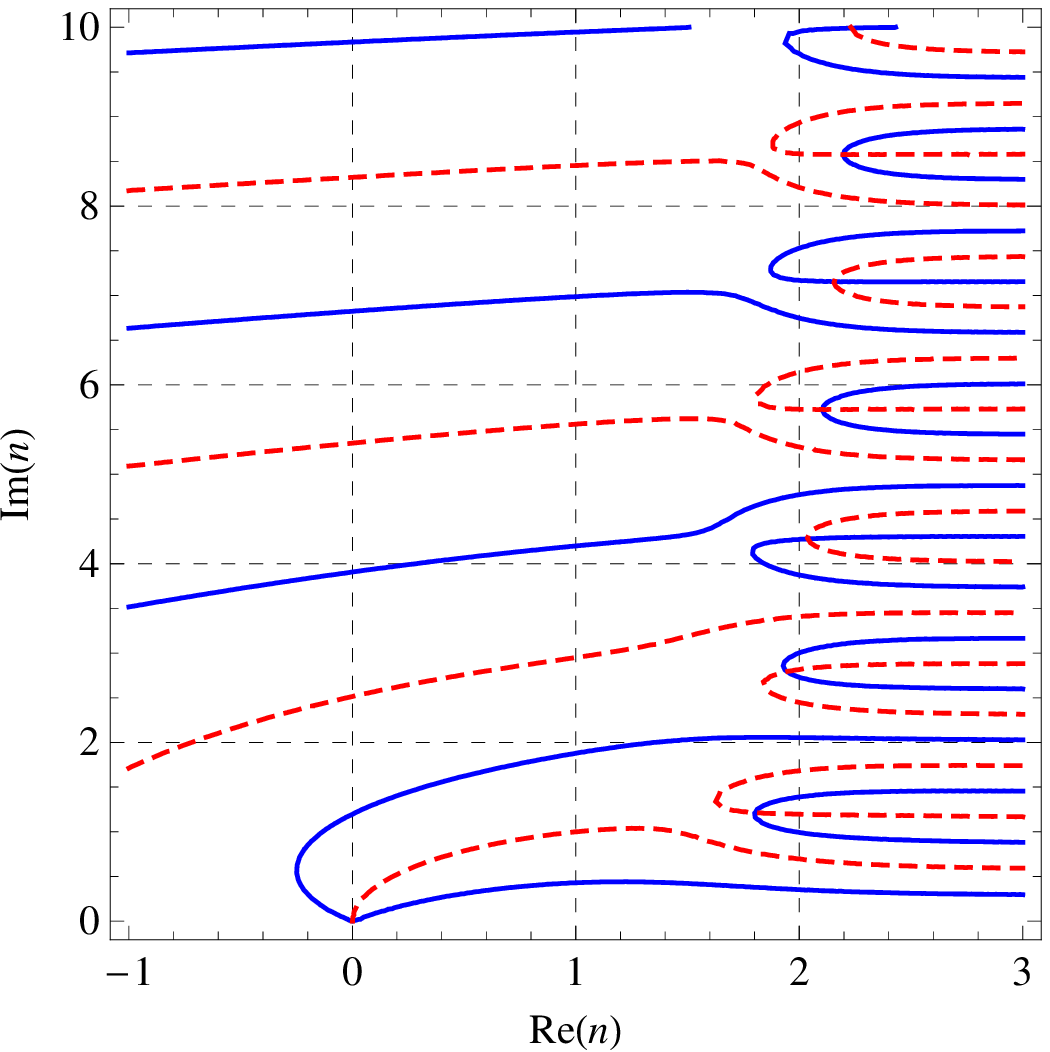}
\caption{}\label{fig5}
\end{figure}

Examples of $\nu$-spectra obtained by solving Eq.~(\ref{consist}) numerically are presented in Table~I. Figure~\ref{fig5} illustrates graphical determination of the lower parts of $n$-spectra. They show that all solutions have positive real parts.

\begin{table}
\begin{tabular}{ccccc}
\hline\hline
\multicolumn{5}{c}{$\nu_m(\theta)$\rule[-4mm]{0mm}{10mm}}\\
\hline\hline
$m$  &  $\theta = 1.5$ & $\theta = 5.5$ & $\theta = 8.5$  & $j2\pi m$ \rule[-4mm]{0mm}{10mm}\\
\hline
$0$&$3.89$&--&--&$0$ \rule{0mm}{6mm}\\
$1$&$3.19+j7.39$&$8.12+j 5.46$&$15.80+j4.17$&$j6.28$ \\
$2$&\rule{7mm}{0mm}$3.75+j13.84$\rule{7mm}{0mm}&\rule{7mm}{0mm}$8.75+j 12.57$\rule{7mm}{0mm}&\rule{7mm}{0mm}$16.31+j11.39$\rule{7mm}{0mm}&\rule{7mm}{0mm}$j12.56$\rule{7mm}{0mm}\\
$3$&$4.11+j20.20$&$9.17+j 19.25$&$16.72+j18.21$&$j18.85$ \\
$4$&$4.37+j26.52$&$9.47+j 25.76$&$17.05+j24.85$&$j25.13$ \\
$5$&$4.58+j32.83$&$9.70+j 32.20$&$17.32+j31.39$&$j31.42$ \\
$6$&$4.76+j39.14$&$9.89+j 38.60$&$17.54+j37.86$&$j37.70$ \\
$7$&$4.90+j45.44$&$10.05+j44.96$&$17.73+j44.29$&$j43.98$ \\
$8$&$5.03+j51.73$&$10.18+j51.31$&$17.89+j50.70$&$j50.27$ \\
$9$&$5.15+j58.02$&$10.30+j57.65$&$18.03+j57.08$&$j56.55$ \\
$10$&$5.23+j64.31$&$10.41+j63.97$&$18.15+j63.45$&$j62.83$\rule[-5mm]{0mm}{5mm}\\
\hline
$\theta\ln\theta$& --& $9.4$ & $18.2$&\rule[-4mm]{0mm}{10mm} \\
\hline\hline
\end{tabular}
\caption{Lower parts of the perturbation growth rate spectra obtained by solving Eq.~(\ref{consist}) numerically. The eigenvalues are measured in units $U_f/b.$ The last row and the last column list their real and imaginary parts as given by the formula (\ref{grate}).}
\label{table2}
\end{table}

To conclude, for sufficiently large values of the incoming fresh-gas
velocity, the piecewise linear V-structure is unstable for
all values of the gas expansion coefficient.

\subsubsection{Space-time profiles of the flow perturbations}

To write down solutions for the flame perturbations, we
need to represent Eq.~(\ref{solution}) in a form suitable for
separating its real part. Using the definitions (\ref{notation}),
(\ref{c3c4}), one has
\begin{eqnarray}&&
h(x) = c_3\left(1 + \alpha x\right)^{\alpha}\int\limits_{\beta}^{1 +
\alpha x}dy y^{-\alpha - 1} e^{n y} = c_3\left(1 + \alpha x
\right)^{\alpha}\left[\int\limits_{\beta}^{1} + \int\limits_{1}^{1 +
\alpha x}\right]dy y^{-\alpha - 1} e^{n y} \nonumber\\ && = c_4
\left(1 + \alpha x\right)^{\alpha} + c_3 \left(x +
\frac{1}{\alpha}\right)^{\alpha}\int\limits_{1/\alpha}^{x +
1/\alpha}dy y^{-\alpha - 1} e^{\nu y}\,,
\end{eqnarray}
\noindent Also, the complex phase of one of the coefficients $c_k$
in the linear problem can be chosen arbitrary. We use this to make
$c_3$ real. Then, writing $\nu = \nu_1 + j\nu_2,$ $c_k =
|c_k|e^{j\varphi_k},$ one combines the formulas (\ref{flowperturb}),
(\ref{ydef}), (\ref{solution}), (\ref{solution1}), and extracts real
parts of the resulting expressions. Thus, we find
\begin{eqnarray}\label{solutionre} \delta u_-(x,t) &&= |c_1|\left[1 - \left(1 + \alpha x\right)^{\alpha}\right]
e^{\nu_1(t-x)}\cos[\nu_2(t-x) + \varphi_1] \nonumber\\&& +
|c_2|xe^{\nu_1(t-x)}\left\{ \nu_1 \cos[\nu_2(t-x) + \varphi_2] -
\nu_2 \sin[\nu_2(t-x) + \varphi_2] \right\}
\nonumber\\&& + c_3\left(x +
\frac{1}{\alpha}\right)^{\alpha}\int\limits_{1/\alpha}^{x +
1/\alpha}dy y^{-\alpha - 1}e^{\nu_1(y + t - x)} \cos[\nu_2(y + t -
x)]\,.
\end{eqnarray}
\noindent The corresponding expression for the $w$-component follows
then from Eq.~(\ref{chuplinr})
\begin{eqnarray}\label{chuplinr2}
\delta w_-(x,t) = \frac{1 - x}{s}\frac{\partial}{\partial x}\delta
u_-(x,t)\,.\end{eqnarray} \noindent Finally, in terms of the
function $g(x),$ the linearized evolution equation
(\ref{evolutionlin}) takes the form
$$\left(e^{\nu x}\tilde{f}(x)\right)' = g(x)[1+\nu(1 - x)] + (x-1)g'(x)\,.$$
Substituting the solution (\ref{solution}), and integrating gives
\begin{eqnarray}\label{evolvg}&&
\tilde{f}(x) = \left[-\frac{\nu}{\alpha + 2}\left(x +
\frac{1}{\alpha}\right)^2 + \left(1 + \frac{\nu}{\alpha}\right)x +
\frac{\nu}{\alpha^2} - 1\right]e^{-\nu x}h(x) + \frac{c_3
e^{\nu/\alpha}}{\alpha + 2}\left(x - \frac{1}{\nu} -
\frac{1}{\alpha} - 1 \right) \nonumber\\&& + x e^{-\nu x}[c_1(1 +
\nu) - c_2\nu] + x^2e^{-\nu x}\nu \left[c_2 \left(1 + \frac{\nu}{2}
\right) - \frac{c_1}{2} \right] - x^3e^{-\nu x}\frac{c_2\nu^2}{3} +
c_5 e^{-\nu x}\,,
\end{eqnarray}
\noindent where $c_5$ is a constant. Its value is fixed by the
condition (\ref{boundarycond3})
$$c_5 = c_1\left[\frac{\nu}{\alpha^2}\frac{\alpha + 1}{\alpha + 2} - 1\right]
+ c_3\frac{e^{\nu/\alpha}}{\alpha + 2}\left(\frac{1}{\nu} +
\frac{1}{\alpha} + 1\right)\,.$$ The perturbed front shape is given
by
$$\delta f(x,t) = {\rm Re}\left\{\tilde{f}(x)e^{\nu t}\right\}\,,$$ which we do
not write out explicitly because of its complexity.

All expressions above are written for $x>0.$ They can be easily
continued to $x<0$ using parity properties of the flow variables.

\section{Discussion and conclusions}\label{conclusions}

The results of analytical investigation presented in this paper give
an accurate and complete account of the stability properties of confined
V-flames anchored in high-velocity streams. The general conclusion we arrived
at is that in this case, the piecewise linear V-structure is unstable for
all values of the gas expansion coefficient. The perturbation growth rate spectra have a similar structure for all $\theta,$ obeying simple classification with respect to the imaginary part of eigenvalues. The only exception is the existence of aperiodic unstable modes for flames with $\theta < \theta_0 \approx 1.8.$ We have found also explicit analytic expressions for the eigenfunctions [Eqs.~(\ref{solutionre})--(\ref{evolvg})].

One result that deserves special emphasis is that dynamics of flame
disturbances in the high-velocity limit turned out to be governed by
the memory effects associated with vorticity generated by the curved
front, which completely dominate contributions due to gas-velocity
jumps across the front that define flame behavior in potential
models. This is in striking contrast with what has been found for
freely propagating flames, where development of the Darrieus-Landau
instability is determined mainly by the structure of these jumps (in
the Sivashinsky-Clavin \citep{sivclav} and Frankel \citep{frankel1990}
models, for instance, memory effects are completely neglected).

Furthermore, dependence of the solution on the gas expansion
coefficient, in particular, appearance of the factors
$\theta^{\theta}$ in Eqs.~(\ref{linsystem1}) -- (\ref{consist1}) is also
quite revealing. It is the result of non-perturbative account of the
influence exerted by the basic flow upon flame disturbances.
Needless to say that such effects cannot be captured in principle by models based on
weak-nonlinearity assumptions.

Our investigation was based on the on-shell description of flames,
developed in Refs.~\citep{kazakov1,kazakov2,jerk1,jerk2}, and extended to the of
anchored flames in Sec.~\ref{anchored}. This formulation allowed us
to elucidate the role of the anchoring system and its influence on
the flame structure, as well as to identify relevant boundary
conditions for the flow variables. The simple and natural way this
analysis was accomplished clearly demonstrates the power of this
approach, not saying about that it permitted the analysis to be
carried out at all.

Another important technical aspect of our work is the locality issue discussed in Secs.~\ref{rodinfluence}, \ref{dconds}, \ref{solutionint}. As we have seen, the requirement of locality of the rod influence on the flame structure appears in the steady case analysis as the consistency condition (\ref{dcond}). On the one hand, this condition expresses the fact that the piecewise constant gas flows of the basic V-pattern satisfy the main integro-differential equation (\ref{generalc1st}), and on the other hand, it serves for selecting inner solutions compatible with the given global flame structure. It is remarkable that the rod influence remains local also in the presence of flame disturbances. Namely, it was proved in Sec.~\ref{analytsol} that jumps in the functions $\tilde{\omega}(x)$ and $E(x)$ at $x=0,$ which are potential sources of nonlocality, vanish in the high-velocity limit.

The last important point to discuss is the practical conditions for
applicability of the results obtained within the large-$U$ limit. As
is evident from the derivations of Sec.~\ref{derivation}, in
practical terms the condition $U\to \infty$ means that $U$ should be
large compared to $(\theta-1).$ At the same time, it is to be noted
that validity of the asymptotic expansion of $\hat{\EuScript{H}},$
obtained in Sec.~\ref{hexpansion}, requires only that $U$ be large
in comparison with unity. The latter condition is considerably
weaker, taking into account that for real flames $\theta$ is
normally $5$ to $8.$ This fact opens a way for investigation of moderate stream-velocities, which is the subject of the subsequent paper \citep{jerk4}.
Another important issue is the influence of gravity. Recent experiments with open flames \citep{cheng1,cheng2} demonstrate that the development of flame disturbances is strongly affected by the gravitational field. This effect can also be studied within our approach.

\acknowledgements{The work presented in this paper was carried out
at the {\it Laboratoire de Combustion et de D\'etonique}. One of the
authors (K.A.K.) thanks the {\it Centre National de la Recherche
Scientifique} for supporting his stay at the Laboratory as a {\it
Chercheur Associ\'e}.}

\begin{appendix}

\section{The Darrieus-Landau relation}\label{appendixA}

In this appendix, we will demonstrate convenience of using two
different imaginary units simultaneously for carrying out actual
calculations. Namely, we will reproduce the classical result of
linear stability analysis for planar flames, which will also serve
as an important check of calculations that led us to
Eq.~(\ref{linearized2}).

In the case of freely propagating planar flames, one has $s=0,$
$U=1,$ $\omega_0 = \theta,$ so that Eq.~(\ref{linearized2})
simplifies to
\begin{eqnarray}\label{linears0}&&
2\tilde{\omega}' + \frac{\theta - 1}{2}\left(1 +
i\hat{H}\right)\left\{\frac{\nu}{\theta}e^{i\nu x/\theta}
\int\limits_{-1}^{+1}d\eta [\tilde{w}(\eta) +
\tilde{f}'(\eta)]\right.\nonumber\\&& \left.\times
e^{-i\nu\eta/\theta} \left[i\cot\left(\frac{\nu}{\theta}\right) +
\chi(x - \eta)\right] - 2i\tilde{f}'(x)
\phantom{\int\limits_{-1}^{+1}}\hspace{-0,5cm} \right\}' = 0\,,
\end{eqnarray}
\noindent where $\hat{H}$ is the ordinary Hilbert operator,
\begin{eqnarray}\label{hilbert}
\hat{H}\exp(i k x) = i\chi(k)\exp(i k x)\,,
\end{eqnarray}\noindent and we took into account the contribution due to
$(x\to -x)^*$ by extending the range of $\eta$-integration and
doubling the last term. The linearized evolution equation takes the
form
\begin{eqnarray}\label{evolutionlins0}&&
\tilde{u}(x) = \nu\tilde{f}(x)\,.
\end{eqnarray}
\noindent As usual, it is most convenient to look for a solution of
these equations in a complex form. In doing so, however, one should
be careful in respecting the original complex structure of
Eq.~(\ref{linears0}). In order to preserve it, one can proceed in
three different ways. The first is to extract the real and imaginary parts of Eq.~(\ref{linears0}), and then proceed to solving the system of equations in the usual way. This is the least convenient means, because it destroys the natural complex structure of Eq.~(\ref{linears0}). Another way followed in Ref.~\citep{jerk2} is to keep all intermediate relations involving the flow variables in an
explicitly real form, like for instance in
Eq.~(\ref{evolutionlins0}). The third method we choose here is to
introduce a new imaginary unit, $j,$ such that $$j^2 = -1, \quad j^*
= j,$$ where the asterisk denotes the complex conjugation with
respect to the initial imaginary unit, $i,$ which had been used in
the derivation of Eq.~(\ref{linears0}),
$$i^* = -i,$$ while the product $(ij)$ is left unspecified. Thus, we
write $$\nu = \nu_1 + j\nu_2, \quad \tilde{u}(x) = \tilde{u}e^{jkx},
\quad \tilde{u} = \tilde{u}_1 + j\tilde{u}_2, \quad {\rm etc.},$$
where $k$ is the wavenumber of perturbation, which according to the
$2$-periodicity condition takes on the values
$$k = \pi m, \quad m\in Z.$$ The physical solution is eventually
found by extracting the real (or imaginary) part of the complex
solution with respect to the unit $j.$

One has
\begin{eqnarray}&&
\int\limits_{-1}^{+1}d\eta e^{jk\eta} e^{-i\nu\eta/\theta}
\left[i\cot\left(\frac{\nu}{\theta}\right) + \chi(x - \eta)\right] =
\frac{1}{jk
-i\nu/\theta}\left\{i\cot\left(\frac{\nu}{\theta}\right)[e^{jk
-i\nu/\theta} - e^{-jk + i\nu/\theta}] \right.\nonumber\\&&\left. +
2 e^{(jk -i\nu/\theta)x} - e^{-jk +i\nu/\theta} - e^{jk
-i\nu/\theta}\right\}  = \frac{2 e^{(jk -i\nu/\theta)x}}{jk
-i\nu/\theta}\,,\nonumber
\end{eqnarray}
\noindent where the constant terms in the curly brackets cancel by
virtue of the condition $e^{2jk}=1.$ Using this in
Eq.~(\ref{linears0}) yields
\begin{eqnarray}
2\tilde{\omega}' + (\theta - 1)\left(1 +
i\hat{H}\right)\left\{\frac{\nu \tilde{w}(x) -
ijk\theta\tilde{f}'(x)}{jk\theta -i\nu}\right\}' = 0\,.\nonumber
\end{eqnarray}
\noindent Multiplying this equation by $(jk\theta - i\nu),$ and
extracting its real (with respect to $i$) part, we find
\begin{eqnarray}\label{linears01}&&
2jk\theta\tilde{u}' + 2\nu\tilde{w}' + (\theta - 1)\left\{\nu
\tilde{w}(x) + jk\theta\hat{H}\tilde{f}'(x) \right\}' = 0\,,
\end{eqnarray}
\noindent while extraction of the imaginary part gives a similar
equation, and comparison of the two leads to the relation
$$\tilde{w}' = \hat{H}\tilde{u}',$$ which can be obtained also
directly from $(1 - i\hat{H})\tilde{\omega}' = 0.$ Finally, writing
$\tilde{u}' = jk\tilde{u},$ $\tilde{f}' = jk\tilde{f},$ and
expressing gas velocity via $\tilde{f}$ with the help of
Eq.~(\ref{evolutionlins0}) leads, after dividing by
$\theta|k|\tilde{f},$ to an algebraic equation
$$\frac{\theta + 1}{\theta}\nu^2 + 2\nu |k| - (\theta - 1)k^2 = 0,$$
from which the well-known Darrieus-Landau dispersion relation for
the perturbation growth rate follows \citep{darrieus,landau}.
$$\nu = \frac{\theta}{\theta + 1} \left(\sqrt{1 + \theta -
\frac{1}{\theta}} \pm 1\right)|k| .$$

\section{Extension of Eq.~(\ref{hcurvedf3}) to discontinuous functions}

If the function $a(x)$ in Eq.~(\ref{hcurvedf3}) does not satisfy conditions
\begin{eqnarray}\label{contcond}
a(0^+) = a(0^-)\,, \quad a(+1) = a(-1)\,,
\end{eqnarray}
\noindent
its derivative is singular at $x=0,\pm 1,$ and the integration by parts used in the transition from Eq.~(\ref{hcurvedf2}) to Eq.~(\ref{hcurvedf3}) is ambiguous. To correctly evaluate the integral, one has to turn back to the exact formula (\ref{hcurvedf}) in which all the functions involved are smooth, and apply it to a function $A(x)$ satisfying (\ref{contcond}), whose behavior near the rod or channel walls looks discontinuous from the outer point of view. More precisely, $A(x)$ is supposed to vary rapidly for $|x| < R \ll 1$ and near the walls, but normally at the intervals $R < x < 1 - R$ and $-1 + R < x < - R,$ where it coincides with $a(x).$ Thus, $$\lim\limits_{R\to 0}A(x) = a(x)\,.$$ We also replace the function $s|x|$ describing the basic V-pattern by a smooth function $F(x)$ such that $$\lim\limits_{R\to 0}F(x) = s|x|\,, \quad x\in (-1,1)\,.$$
Neglecting the anchor dimensions means that the action of $\hat{\EuScript{H}}$ on $a'$ is defined as
$$\left(\hat{\EuScript{H}}a'\right)(x) = \lim\limits_{R\to 0}\left\{\left(\hat{\EuScript{H}}A'\right)\right\}(x)\,.$$
To find out how $\hat{\EuScript{H}}$ acts on the derivative of $A(x),$ we replace $a$ by $A$ in Eq.~(\ref{hcurvedf2}), and integrate the right hand side by parts
\begin{eqnarray}\label{hcurvedfa1}
\left(\hat{\EuScript{H}}A'\right)(x) &=& \frac{1 + i
F'(x)}{2}~\fint\limits_{-1}^{+1}
d\eta~A'(\eta)\cot\left\{\frac{\pi}{2}(\eta - x +
i[F(\eta) - F(x)])\right\} \nonumber\\ &=& \frac{1}{2}\frac{d}{dx}\fint\limits_{-1}^{+1}
d\eta~[1 + iF'(\eta)]A(\eta)\cot\left\{\frac{\pi}{2}(\eta - x +
i[F(\eta) - F(x)])\right\}\,.
\end{eqnarray}
\noindent The boundary terms vanish here because the integral kernel is $2$-periodic, and $A(x)$ satisfies $A(-1)=A(+1),$ by the assumption. Since the functions $A(x)$ and $F'(x)$ have only finite jumps in the limit $R \to 0,$ the last integral in Eq.~(\ref{hcurvedfa1}) is well-defined in this limit, representing a continuously differentiable function for all $|x|\in (0,1).$  Thus,
$$\lim\limits_{R\to 0}\left\{\left(\hat{\EuScript{H}}A'\right)\right\}(x) = \frac{1}{2}\frac{d}{dx}\fint\limits_{-1}^{+1}
d\eta~[1 + is\chi(\eta)]a(\eta)\cot\left\{\frac{\pi}{2}[\eta - x +
is(|\eta| - |x|)]\right\}\,.$$

Next, we go over to the large-slope limit. The right hand side of the last equation can be evaluated in this case in exactly the same way as we arrived to Eq.~(\ref{hcurvedf2}). Comparison with Eq.~(\ref{intred}) shows that the role of the function $a(\eta)$ in this equation is now played by $[1 + is\chi(\eta)] a(\eta),$ the only difference being that the large factor $s$ comes from the integrand, rather than from the pre-integral factor in Eq.~(\ref{hcurvedf}). Taking this into account, we readily find
\begin{eqnarray}&&
\left(\hat{\EuScript{H}}a'\right)(x) = \frac{1}{2}\frac{d}{dx}\left[\int\limits_{0}^{1}d\eta \left\{a(\eta)[s\chi(\eta) - i] + a(-\eta)[s\chi(-\eta) - i]\right\}\chi(\eta - |x|) \right.\nonumber\\&& \left.- ia(-x)(2|x| - 1)\phantom{\int}\hspace{-0,4cm}\right] = - s\chi(x)\left\{a(|x|) - a(-|x|)\right\} + i\chi(x)\left\{a(|x|) + a(-|x|)\right\} \nonumber\\&& - 2ia(-x)\chi(x) + ia'(-x)(2|x| - 1)\,. \nonumber
\end{eqnarray}
\noindent Using the obvious identity $\chi(x)\{a(|x|) + a(-|x|) - 2a(-x) \} = a(|x|) - a(-|x|),$ we finally obtain
\begin{eqnarray}&&
 \left(\hat{\EuScript{H}}a'\right)(x) =
 (s\chi(x) - i)\left\{a(-|x|) - a(|x|)\right\} + ia'(-x)(2|x| - 1)\,, \nonumber
\end{eqnarray}
\noindent which is exactly Eq.~(\ref{hcurvedf3}), as was to be proved. Note that this result is independent of the particular choice of the functions $A(x),F(x).$

\end{appendix}

\bibliography{references}
~\newpage

~\\\\
{\large\bf List of figures}\\\\ Channel propagation of a flame
anchored by a cylindrical rod of radius $R,$ located downstream.
\dotfill 23\\ Geometry of the integrand in expression
(\ref{integral}) in the case $n=1,$ $\eta \in [0,b].$ \dotfill 24\\
Contour deformation used to calculate the integral on the right of
Eq.~(\ref{yintegral}). The initial and deformed contours are shown
by the full and broken lines, respectively. The crosses denote poles
of the hyperbolic cotangent. \dotfill 25\\
The coefficient $S$ in Eq.~(\ref{consist1}) versus gas expansion coefficient (solid line). Broken line is the function $\theta^{\theta}.$ \dotfill 26\\
Curves representing the real (solid lines) and imaginary (dashed lines) parts of Eq.~(\ref{consist}) for $\theta = 5.5.$ The roots correspond to the lines intersections.\dotfill 27\\

~\newpage
~\newpage
~\newpage
~\newpage
~\newpage

\end{document}